\documentclass[%
 reprint,
superscriptaddress,
 amsmath,amssymb,
 aps,
]{revtex4-2}

\usepackage{graphicx}
\usepackage{dcolumn}
\usepackage{bm}
\usepackage{siunitx}

\usepackage{booktabs}

\begin{document}

\preprint{APS/123-QED}

\title{Red Emission from Copper-Vacancy Color Centers in Zinc Sulfide Colloidal Nanocrystals}

\author{Sarah M. Thompson}
\affiliation{Department of Electrical and Systems Engineering, University of Pennsylvania, Philadelphia Pennsylvania 19104, USA}
\author{C\"uneyt \c{S}ahin}
\affiliation{UNAM~--~National Nanotechnology Research Center and Institute of Materials Science and Nanotechnology, Bilkent University, Ankara, Turkey}
\affiliation{Department of Physics and Astronomy, University of Iowa, Iowa City IA, 52242, USA}
\author{Shengsong Yang}
\affiliation{Department of Chemistry, University of Pennsylvania, Philadelphia PA, 19104, USA}
\author{Michael E. Flatt\'e}
\affiliation{Department of Physics and Astronomy, University of Iowa, Iowa City IA, 52242, USA}
\affiliation{Department of Applied Physics, Eindhoven University of Technology, P. O. Box 513, 5600 MB Eindhoven, The Netherlands}
\author{Christopher B. Murray}
\affiliation{Department of Chemistry, University of Pennsylvania, Philadelphia PA, 19104, USA}
\affiliation{Department of Materials Science and Engineering, University of Pennsylvania, Philadelphia PA, 19104, USA}
\author{Lee C. Bassett}
\email[Corresponding authors. ]{lbassett@seas.upenn.edu \& kagan@seas.upenn.edu }
\affiliation{Department of Electrical and Systems Engineering, University of Pennsylvania, Philadelphia PA, 19104, USA}
\author{Cherie R. Kagan}
\email[Corresponding authors. ]{lbassett@seas.upenn.edu \& kagan@seas.upenn.edu }
\affiliation{Department of Electrical and Systems Engineering, University of Pennsylvania, Philadelphia Pennsylvania 19104, USA}
\affiliation{Department of Materials Science and Engineering, University of Pennsylvania, Philadelphia Pennsylvania 19104, USA}
\affiliation{Department of Chemistry, University of Pennsylvania, Philadelphia Pennsylvania 19104, USA}

\date{\today}

\begin{abstract}
Copper-doped zinc sulfide (ZnS:Cu) exhibits down-conversion luminescence in the UV, visible, and IR regions of the electromagnetic spectrum; the visible red, green, and blue emission is referred to as R-Cu, G-Cu, and B-Cu, respectively. 
The sub-bandgap emission arises from optical transitions between localized electronic states created by point defects, making ZnS:Cu a prolific phosphor material and an intriguing candidate material for quantum information science, where point defects excel as single-photon sources and spin qubits. 
Colloidal nanocrystals (NCs) of ZnS:Cu are particularly interesting as hosts for the creation, isolation, and measurement of quantum defects, since their size, composition, and surface chemistry can be precisely tailored for bio-sensing and opto-electronic applications.
Here, we present a method for synthesizing colloidal ZnS:Cu NCs that emit primarily R-Cu, which has been proposed to arise from the Cu\textsubscript{Zn}-V\textsubscript{S} complex, an impurity-vacancy point defect structure analogous to well-known quantum defects in other materials that produce favorable optical and spin dynamics.
First principles calculations confirm the thermodynamic stability and electronic structure of  Cu\textsubscript{Zn}-V\textsubscript{S}.
Temperature- and time-dependent optical properties of ZnS:Cu NCs show blueshifting luminescence and an anomalous plateau in the intensity dependence as temperature is increased from 19 K to 290 K, for which we propose an empirical dynamical model based on thermally-activated coupling between two manifolds of states inside the ZnS bandgap. 
Understanding of R-Cu emission dynamics, combined with a controlled synthesis method for obtaining R-Cu centers in colloidal NC hosts, will greatly facilitate the development of Cu\textsubscript{Zn}-V\textsubscript{S} and related complexes as quantum point defects in ZnS.   
\end{abstract}

\maketitle{}

\section{\label{sec:level1}Introduction}
Controlled impurity doping of wide-bandgap semiconductors can be used to introduce color centers, which are point defects that activate sub-bandgap, optical photoluminescence (PL). 
Color centers can serve as sources of tunable PL for bio-imaging and opto-electronic applications\cite{Castelletto2021ColorReview, Lahariya2022}, as well as localized, optically-addressable, electronic spin states for applications in quantum information science\cite{Norman2021,Bassett2019}. 
For all of these applications, colloidal nanocrystals (NCs) can provide unique advantages over analogous, bulk materials because they can be processed using wet-chemical methods, and their large surface areas and effects of quantum confinement allow for highly tunable optical and electronic properties\cite{Kagan2020ColloidalScience}.
While the focus of PL studies in impurity-doped NCs is frequently on emission mechanisms that involve the host bandgap and confinement effects, control of deeply sub-bandgap PL emission associated with localized defect states is important for developing suitable color centers in new materials for QIS applications.

Impurity-doped ZnS has long been exploited as a UV, visible, and NIR luminescent material in its bulk and colloidal NC forms, and it has more recently been studied as a potential host material for defect-based quantum emitters and quantum bits, or defect qubits\cite{Stewart2019QuantumSulfide, Phys2022}. 
Cu-doping of ZnS introduces sub-bandgap red, green, and blue PL emission bands that are associated with color centers known respectively as R-Cu, G-Cu, and B-Cu. 
R-Cu color centers are particularly interesting thanks to their peak PL emission in the NIR biological window.
However, R-Cu remains under-utilized since it is rarely observed in colloidal ZnS:Cu NCs, which typically emit visible PL dominated by B-Cu and G-Cu\cite{Knowles2016, Bol2002, Curcio2019}.

Past studies have indicated that R-Cu emission in bulk ZnS:Cu arises from a defect complex consisting of a substitutional copper impurity (Cu\textsubscript{Zn}) and a sulfur vacancy (V\textsubscript{S}) in a nearest-neighbor Cu\textsubscript{Zn}-V\textsubscript{S} complex\cite{Shionoya1966NatureMeasurements}. 
Compared to transition metal-doped phosphors like ZnS:Mn that rely on electric-dipole-forbidden, intra-$d$-shell transitions between substitutional Mn\textsubscript{Zn} levels to produce visible PL, the mixed orbital character and lowered symmetry of the Cu\textsubscript{Zn}-V\textsubscript{S} complex are associated with more dipole-allowed radiative transitions\cite{Yen2007}, and therefore shorter optical lifetimes, as desired for many applications.
Moreover, the symmetry of the Cu\textsubscript{Zn}-V\textsubscript{S} complex is described by the $C_{3v}$ point group, which is characteristic of well-developed defect qubits \cite{Doherty2011, Koehl2011} and is a key factor in producing favorable defect orbital and spin structures\cite{Bassett2019}.  
R-Cu emission has further been associated with electron paramagnetic resonance (EPR) spectra which indicate a paramagnetic ground state \cite{Holton1969ParamagneticZnS}. 
These characteristics are especially compelling in combination with the favorable properties of ZnS as a host for defect qubits, which include a high natural abundance of spin-free nuclei and relatively weak spin-orbit coupling, as well as the ease of ZnS colloidal NC synthesis and surface modification compared to hosts materials such as diamond \cite{Kagan2020ColloidalScience, Wolfowicz2021}.

Here, we report the synthesis and characterization of colloidal ZnS:Cu NCs emitting visible PL dominated by R-Cu. 
We study the structural, compositional, and time- and temperature-dependent optical properties of NCs synthesized with 0, 0.05, 0.075, and 0.1 mol\% Cu:Zn. 
The R-Cu emission intensity scales with the copper concentration, and the R-Cu emission band exhibits complex temperature- and time-dependent properties.
In particular, the R-Cu emission peak blueshifts with increasing temperature from 19 K to 290 K, consistent with observations in bulk ZnS:Cu\cite{Shionoya1966NatureMeasurements}, and the R-Cu emission intensity as a function of temperature, of which there are no reports from bulk ZnS:Cu to our knowledge, shows an anomalous plateau between 150 K and 270 K. 
We propose a single mechanism to explain the temperature-dependent peak energy, intensity, and lifetime based on thermally-activated carrier transfer between two manifolds of radiative states. 
This mechanism is consistent with time-resolved PL measurements showing the presence of two distinct radiative transitions in the R-Cu band at low temperature. 
Drawing from first-principles calculations, we discuss the role of defect species, spatial arrangement, and charge state in producing the manifolds of states responsible for measured R-Cu characteristics. 
A detailed understanding of these characteristics can facilitate the realization of protocols for initialization, readout, and control of the defect's charge and spin states for development of a quantum defect architecture. 

\section{Results and Discussion}

\subsection{Synthesis of ZnS:Cu NCs with R-Cu Emission}
\begin{figure}
\includegraphics[scale=0.95]{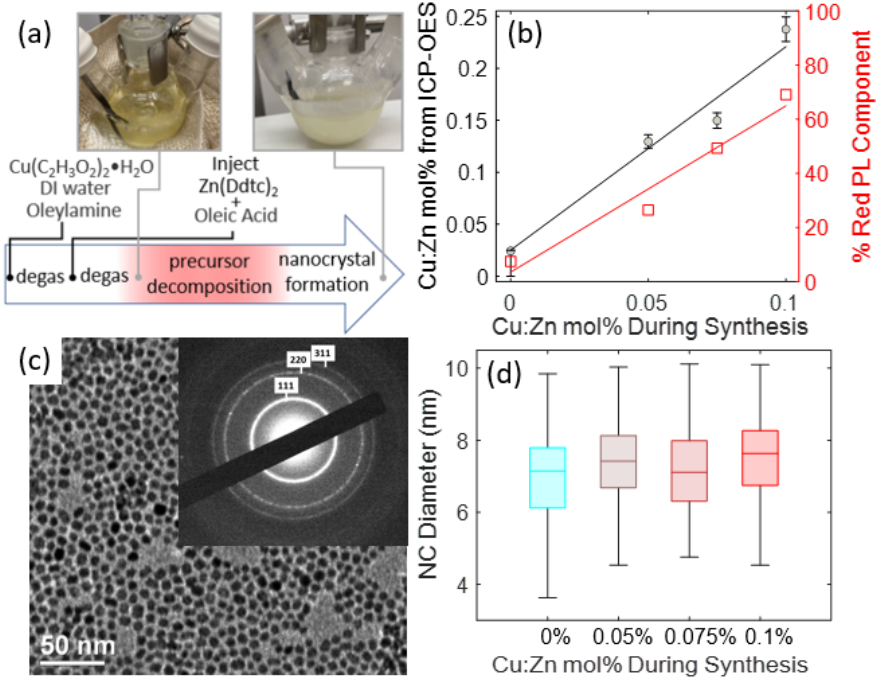}
\caption{\label{fig:Synthesis and Structure}
 \textbf{ZnS:Cu NC synthesis and structure.} 
 (a) Schematic of the synthesis of R-Cu emitting ZnS:Cu NCs, where red represents the application of heat. Photographs show the reaction vessel before and after NC formation. 
 (b) Cu:Zn mol\% measured by ICP-OES (black circles) as a function of the Cu:Zn mol\% added to the synthesis pot. 
 The component weight of the R-Cu peak when the total PL spectrum is decomposed by non-negative matrix factorization (red squares and right-hand vertical axis) also scales with the Cu:Zn mol\% . The R\textsuperscript{2} values for the linear regression fits (black and red lines) are 0.917 and 0.957, respectively. 
 (c) TEM image and electron diffraction pattern of a sample of ZnS:Cu NCs with 0.1 mol\% Cu:Zn. 
 (d) Distribution of NC diameters for samples of 100 NCs measured from TEM images obtained for each Cu:Zn ratio (0-0.1 mol\%). 
}
\end{figure}

ZnS NCs are synthesized using the single-source precursor approach developed by Zhang \textit{et al}., \cite{Zhang2010} in which zinc diethyldithiocarbamate (Zn(Ddtc)\textsubscript{2}) is thermally decomposed in oleic acid (OA) and oleylamine (OM); see Figure \ref{fig:Synthesis and Structure}a. 
In previously reported syntheses of colloidal ZnS:Cu NCs, the absence of R-Cu emission may result from unintentional Cl impurities introduced by CuCl\textsubscript{2} precursors, which are known to quench R-Cu in bulk ZnS:Cu along with Al, In, and Ga impurities \cite{Shionoya1966NatureMeasurements,Holton1969ParamagneticZnS,Shionoya1964NatureCrystals}. 
To avoid the introduction of Cl impurities, we instead add a fixed volume (0.1 mL) of Cu(CH\textsubscript{3}COO)\textsubscript{2}$\cdot$H\textsubscript{2}O dissolved in ultrapure DI water, with concentrations corresponding to Cu:Zn molar ratios of 0 \%, 0.05 \%, 0.075 \%, and 0.1 \%, to the synthesis pot prior to degassing. 
In the case of undoped ZnS NCs, the 0.1 mL addition consists of DI water only. 
Inductively-coupled plasma - optical emission spectroscopy (ICP-OES) and PL measurement results (Figure \ref{fig:Synthesis and Structure}b) show that varying the concentration of the Cu precursor directly influences the final Cu concentration in the NC samples, as well as the relative intensity of the red PL emitted by the NCs. 
The PL measurement results are discussed further in the next subsection.

A representative TEM image of ZnS:Cu NCs with 0.1 mol\% Cu:Zn  (Figure \ref{fig:Synthesis and Structure}c) shows that the samples are composed of 7.2$\pm$1.2 nm diameter particles. 
The NC size distribution remains consistent across differently-doped samples (Figure \ref{fig:Synthesis and Structure}d). 
Electron diffraction measurements (inset, Figure \ref{fig:Synthesis and Structure}c) exhibit peaks at 2$\Theta$ values that correspond to the $\langle$111$\rangle$, $\langle$220$\rangle$, and $\langle$311$\rangle$ planes of sphalerite (zinc blende), according to PDF\# 98-000-04053.

\begin{figure}
\includegraphics[scale=0.85]{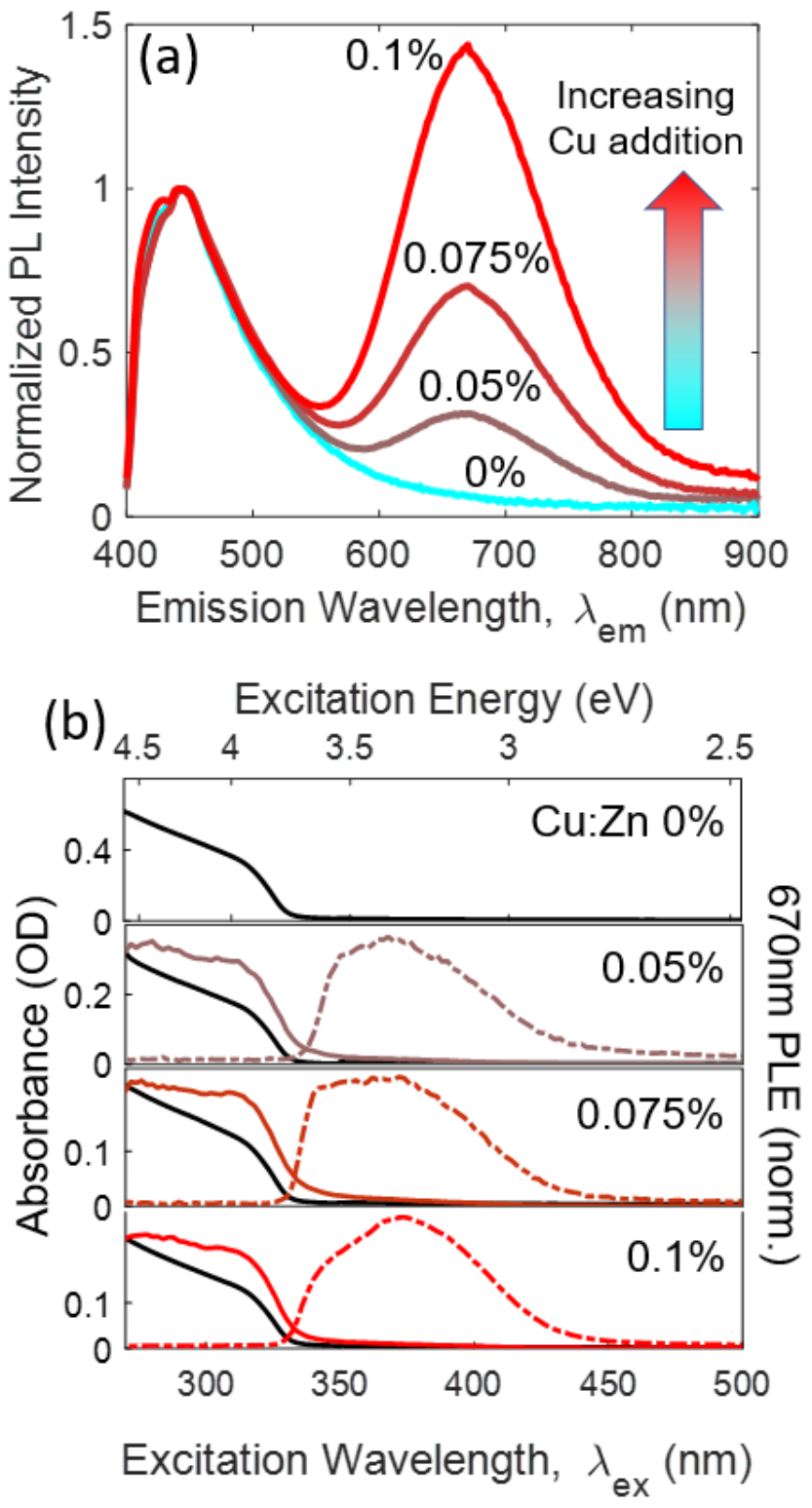}
\caption{\label{fig:RT Optical Characterization}
 \textbf{Room-temperature optical properties.} 
 (a) PL spectra under continuous-wave excitation at $\lambda_\mathrm{ex}$=375 nm, normalized to the PL intensity at $\lambda_\mathrm{em}$=442 nm from ZnS NCs synthesized with 0--0.1 mol\% Cu:Zn. 
 (b) Absorption spectra (black curves) and corresponding $\lambda_\mathrm{em}$=670 nm PLE spectra for dilute (\textless0.5mg/mL) NC dispersions (solid colored curves). Dashed colored curves indicate PLE spectra for concentrated (10mg/mL) NC dispersions with above-bandgap absorbance greater than OD=4.0. PLE spectra monitor the emission intensity at $\lambda_\mathrm{em}$=670 nm as a function of $\lambda_\mathrm{ex}$ and are normalized to their peak wavelength values.
}
\end{figure}

\subsection{Room-Temperature Optical Characterization}

PL emission spectra from NC samples containing the four different Cu concentrations are shown in Figure \ref{fig:RT Optical Characterization}a. 
The intrinsic, background PL peak with emission wavelength, $\lambda_\mathrm{em}$, between 430 nm and 600 nm, is present regardless of Cu concentration. 
This PL feature is characteristic of undoped ZnS NCs and is widely accepted to arise from radiative transitions between intrinsic defect states inside the ZnS bandgap created by Zn and S vacancies (V\textsubscript{Zn} and V\textsubscript{S}) and interstitials (Zn\textsubscript{i} and S\textsubscript{i})\cite{Denzler1998, Goswami2007PhotoluminescentWires, Ramasamy2012SynthesisNanoparticles}. 
Similar PL is also observed from bulk, undoped ZnS, with features being attributed to both intrinsic defects and unintentional impurities\cite{Saleh2019}.
 
Distinct emission with $\lambda_\mathrm{em}$ centered at 670 nm is observed in the Cu-doped NCs with a relative intensity that increases with the Cu:Zn molar ratio. 
The PL spectra of Figure \ref{fig:RT Optical Characterization}a are decomposed using nonnegative matrix factorization (SI Section 1) in order to calculate the relative strengths of the intrinsic and dopant-induced components, yielding the concentration dependence plotted in Figure \ref{fig:Synthesis and Structure}b. 
Absorption spectra and $\lambda_\mathrm{em}$=670 nm PLE spectra for all NC materials are shown in Figure \ref{fig:RT Optical Characterization}b. 
From the absorption spectra, we construct Tauc plots (SI Section 2) and extract bandgap energies between 3.77 eV and 3.79 eV. 

The $\lambda_\mathrm{em}$=670 nm PL and PLE spectra in Figure \ref{fig:RT Optical Characterization} align well with those reported for R-Cu in bulk ZnS:Cu, which also peaks at $\lambda_\mathrm{em}$=670 nm at room temperature and is attributed to transitions between V\textsubscript{S} levels and Cu\textsubscript{Zn} levels inside the ZnS bandgap.\cite{Shionoya1964NatureCrystals} 
The PLE spectra of Figure \ref{fig:RT Optical Characterization}b show that the $\lambda_\mathrm{em}$=670 nm PL in all doped NC samples is excited by the ZnS host, more easily observable in measurements of dilute NC dispersions (\textless0.5mg/mL, solid curves in Figure \ref{fig:RT Optical Characterization}b), as well as by sub-bandgap wavelengths in the range $\lambda_\mathrm{ex}$= 330 -- 450 nm, more readily seen in concentrated (10mg/mL, dashed curves) NC dispersions.
The sub-bandgap excitation allows excitation of a greater number of color centers in concentrated samples because it is more efficiently transmitted throughout the sample volume (SI Section 3). 
Thus, the PLE from concentrated NC dispersions more closely resembles PLE for defect emission from ZnS bulk crystals and powders\cite{Saleh2019, Shionoya1966NatureMeasurements, Bol2002}.
We use sub-bandgap, 375 nm excitation in this work to achieve overall higher PL count rates and allow efficient measurement of temperature-dependent PL from deposited films of ZnS:Cu NCs.
The polarization dependence of the sub-bandgap PLE in bulk ZnS:Cu indicates a nearest-neighbor configuration of V\textsubscript{S} and Cu\textsubscript{Zn} with $C_{3v}$ point-group symmetry\cite{Shionoya1966NatureMeasurements}, consistent with more recent EXAFS measurements in ZnS:Cu NCs\cite{Car2011NanoscaleEXAFS}. 
R-Cu is quenched when bulk ZnS:Cu phosphors are fired in atmospheres containing high sulfur pressure\cite{Shionoya1966NatureMeasurements}, further supporting the role of V\textsubscript{S} levels in the PL. 

Compared to their bulk counterparts, impurity doping of NC materials can be challenging to achieve and to confirm.\cite{Erwin2005}
The alignment between the R-Cu PL/PLE spectra we measure here and those arising from bulk ZnS:Cu is suggestive of successful Cu doping of the ZnS:Cu NCs, since there is extensive evidence that R-Cu in bulk ZnS:Cu is activated by Cu substitutionally occupying Zn sites. 
We additionally carry out studies in which we deposit NC thin films and treat them with methanol and methanolic Na\textsubscript{2}S and Zn(CO\textsubscript{2}CH\textsubscript{3})\textsubscript{2}$\cdot$2H\textsubscript{2}O solutions known to remove organic ligands and to strip surface cations\cite{Goodwin2014}, and to enrich the NC surface in S\textsuperscript{2-} or Zn\textsuperscript{2+}, respectively,\cite{Oh2014, Kim2013} altering the presence or environment of Cu cations if they are on the surface. 
In all cases, the surface treatments do not quench or enhance the R-Cu PL from our NCs, again consistent with successful Cu doping of their cores (SI Section 4).

\subsection{Variable Temperature Studies Probing the Origins of Cu-Induced Sub-Bandgap PL/PLE}
\begin{figure}
\centering
\includegraphics[width=0.5\textwidth]{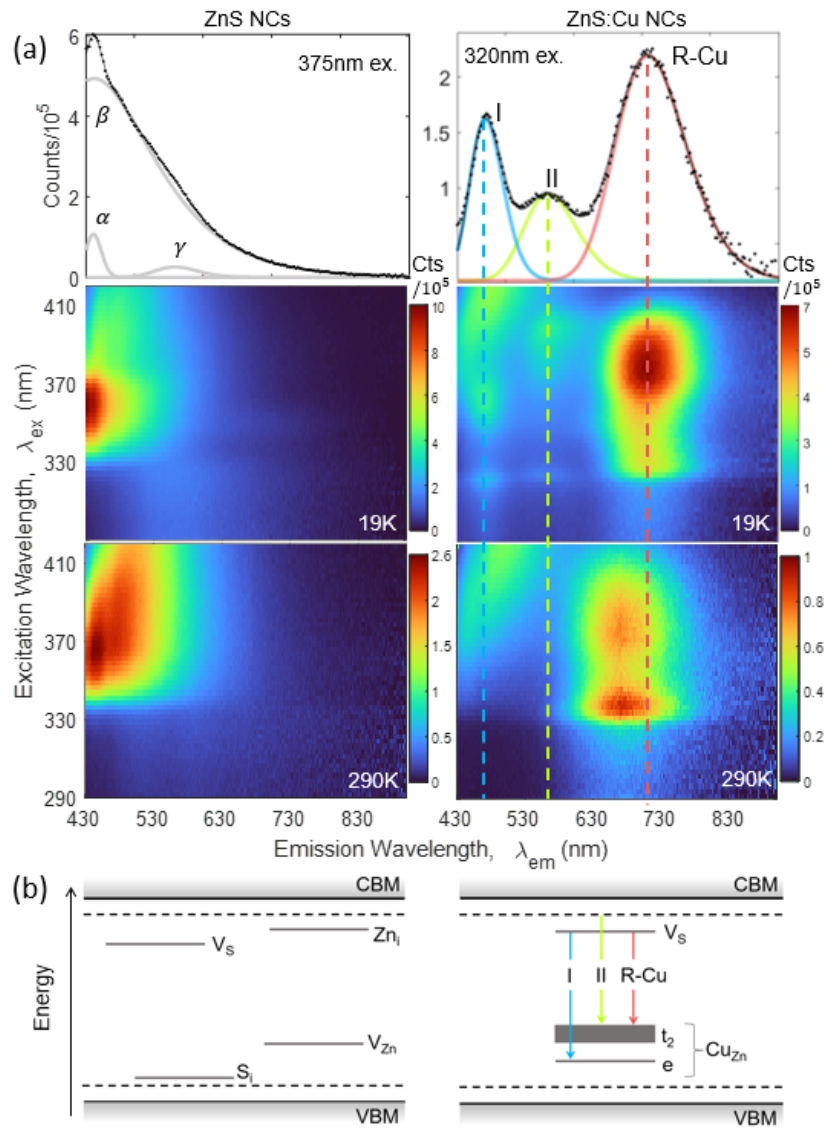}
\centering
\caption{\label{fig:LT PL/PLE}
\textbf{Temperature-dependent PL/PLE}
(a) PL spectra (top) and PL/PLE maps (below) of films of ZnS NCs synthesized with 0 mol\% (left) and 0.1 mol\% Cu:Zn (right), measured at 19 K and room temperature. 
The PL spectra extracted from the 19 K PL/PLE maps are photoexcited at $\lambda_\mathrm{ex}$=375 nm and $\lambda_\mathrm{ex}$=320 nm for the ZnS NC and ZnS:Cu NC films, respectively, to show the clearest peak separation and spectrally reduce intrinsic PL in the case of ZnS:Cu NCs. 
(b) Energy level diagrams showing key defect states responsible for sub-bandgap PL emission in the undoped and doped NCs. Dashed lines represent shallow surface defect states. Arrows are used to indicate assigned radiative transitions in the doped NCs, which involve different sub-levels of the Cu\textsubscript{Zn} 3$d$ shell. The $t_2$ level is indicated with a heavier line to suggest that it may be further split into $e$ and $a$ sub-levels depending on the Cu\textsubscript{Zn} impurity site coordination.
}
\end{figure}

NCs are drop-cast onto sapphire substrates and loaded inside an evacuated cryostat for temperature- and time-dependent spectroscopic measurements. 
Figure \ref{fig:LT PL/PLE} shows PL/PLE maps of ZnS NCs without Cu doping (Cu:Zn at 0 mol\%) and with Cu doping (Cu:Zn  0.1 mol\%), measured at 19 K and at 290 K. 
The NC films are highly absorbing and therefore the above-bandgap PLE is relatively diminished as discussed in the previous section, consistent with the dashed PLE curves in Figure \ref{fig:RT Optical Characterization}b.
PL from the undoped ZnS NCs is dominated by intrinsic PL at all temperatures. The 19 K PL spectrum from the undoped ZnS NCs ($\lambda_\mathrm{ex}$=375 nm) can be described using three Gaussian peaks, which are labeled $\alpha$, $\beta$, and $\gamma$ in Figure \ref{fig:LT PL/PLE}a. 
Peaks $\alpha$ and $\beta$ dominate the PL for $\lambda_\mathrm{ex}>330$ nm (corresponding to sub-bandgap excitation of the ZnS NCs), and their peak emission wavelength varies with $\lambda_\mathrm{ex}$. 
The third peak observable in the undoped film, peak $\gamma$, remains relatively fixed regardless of $\lambda_\mathrm{ex}$ and is the only peak observed in this spectral region for $\lambda_\mathrm{ex}<$330 nm.

For ZnS:Cu NCs, the 19 K PL spectrum ($\lambda_\mathrm{ex}$=320 nm) shows R-Cu emission, as well as blue and green emission peaks labeled I and II (Figure \ref{fig:LT PL/PLE}). 
The three PL peaks are observed for $\lambda_\mathrm{ex}$ ranging from 290 -- 420 nm. 
For $\lambda_\mathrm{ex}<330$ nm, to the blue of the ZnS bandgap wavelength, the sub-bandgap intrinsic PL is significantly diminished in intensity compared to peaks I, II, and R-Cu. 
The R-Cu peak is distinguishable at all temperatures, and the peak emission wavelength blueshifts from 709 nm at 19 K to 670 nm at room temperature.
This observation is similar to the reported blueshift in bulk ZnS:Cu from 700 nm at 4 K to 670 nm at room temperature.\cite{Shionoya1964NatureCrystals,Shionoya1966NatureMeasurements} 
Peaks I and II, with $\lambda_\mathrm{em}$= 471 nm and $\lambda_\mathrm{em}$= 562 nm, respectively, are quenched at room temperature. 
The $\lambda_\mathrm{em}$ and FWHM values for PL labeled in Figure \ref{fig:LT PL/PLE} are listed in Table \ref{table:peaks}.
Line plots of the spectral data in the PL/PLE maps of Figure \ref{fig:LT PL/PLE}a are included in SI Section 5. 

\begin{table*}[t]
\centering
\caption{Peak positions and widths for I, II, R-Cu, and $\alpha$, $\beta$, and $\gamma$, from Gaussian fitting of PL data shown in Fig. \ref{fig:LT PL/PLE}         \label{table:peaks}}
\begin{tabular}{ccccc}
\hline
Cu:Zn mol\% & Label & $\lambda_\mathrm{em}$ (nm) & $E_\mathrm{em}$ (eV) & FWHM (eV)\\   \hline
 0.1&I   &  471 & 2.61  & 0.13 \\
 0.1&II   &  562 & 2.19  & 0.21  \\  
 0.1 &R-Cu   & 709 & 1.74  & 0.27   \\    
 \hline
 0 &$\alpha$& 440 & 2.81 & 0.17   \\
0 &$\beta$ &  442 & 2.65  & 1.12   \\    
0&$\gamma$ & 560 & 2.20 & 0.30   \\
 \hline
\end{tabular} \\
\end{table*}

Figure \ref{fig:LT PL/PLE}b shows energy level diagrams containing key defect levels believed to activate PL in the undoped and doped NCs. 
In undoped ZnS, intrinsic PL is assigned to transitions between Zn\textsubscript{i}, V\textsubscript{S}, V\textsubscript{Zn}, and S\textsubscript{i} levels, for which the relative energy levels shown are agreed upon, but the exact energies are not known\cite{Denzler1998, Kripal2010}. 
PL peaks activated by Cu doping the ZnS NCs, which can be spectrally separated from intrinsic PL as discussed in this section, are assigned to radiative transitions in the diagram. 
R-Cu emission arises from transitions between states primarily associated with V\textsubscript{S} and the Cu\textsubscript{Zn} $t_2$ levels\cite{Shionoya1964NatureCrystals}. 
We assign peak I to a transition between V\textsubscript{S} levels and the lower-lying Cu\textsubscript{Zn} $e$ levels \cite{Bol2002}. 
This assignment is supported by our measurement of a 0.87 eV energy difference between peak I and R-Cu PL, similar to the reported 0.86 eV energy difference between Cu\textsubscript{Zn} $t_2$ and $e$ levels\cite{Peka1994EmpiricalCdS}. 
Peak II is assigned to transitions between donor levels that are shallower than V\textsubscript{S}, attributed here to surface defects, and the Cu\textsubscript{Zn} $t_2$ levels\cite{Xue-Ying2015}. 
We note that the state labels and identifications in Figure \ref{fig:LT PL/PLE}b are based on an approximate picture of isolated V\textsubscript{S} and Cu\textsubscript{Zn} in cubic ZnS, whereas the Cu\textsubscript{Zn}-V\textsubscript{S} is characterized by lowered $C_{3v}$ symmetry and hybridization between these levels.
We discuss this point in more detail later in the next section, drawing insight from theoretical calculations.

\begin{figure}
\centering
\includegraphics[scale=.85]{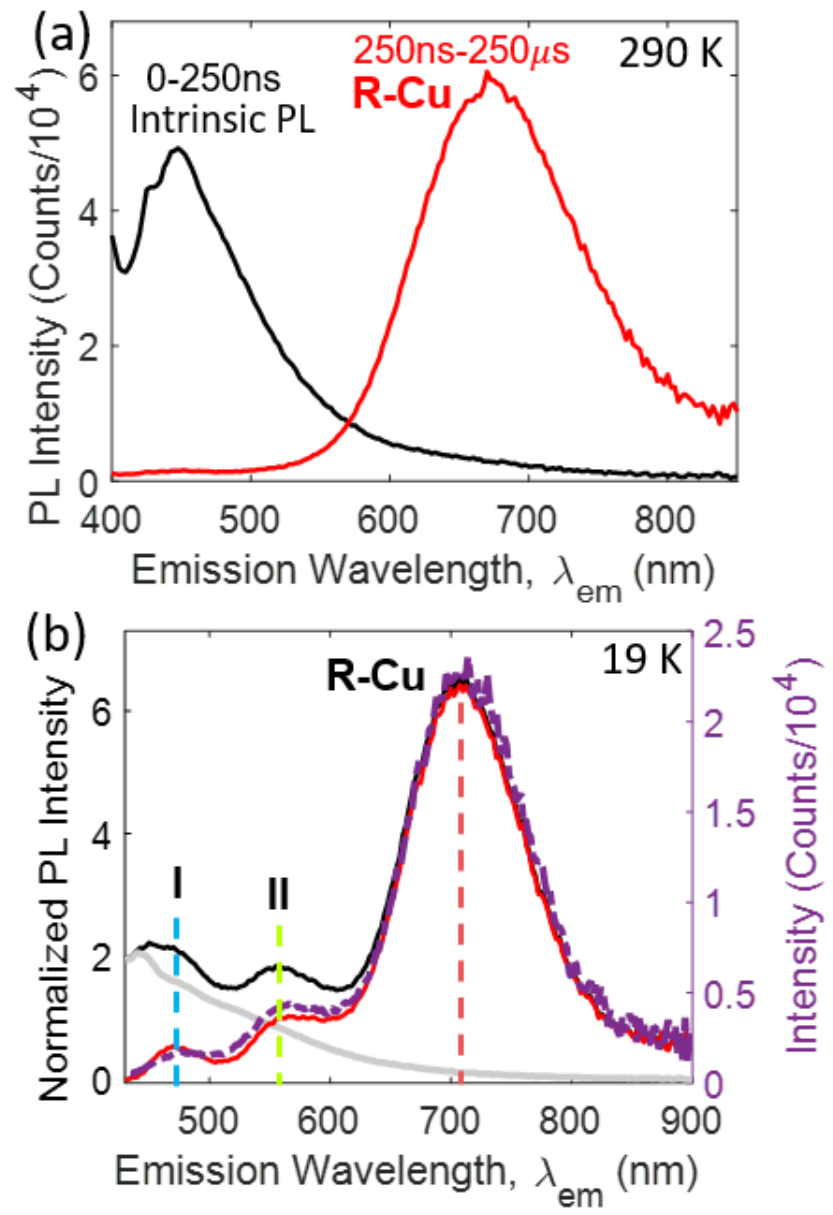}
\centering
\caption{\label{fig:TRES}
\textbf{Isolation of Cu-Activated PL} (a) Time-resolved emission spectra from ZnS:Cu NC films at 290 K under 375 nm, 1 kHz pulsed excitation, in which counts from the first 250 ns following the laser pulse (black) are plotted separately from subsequent counts (red), effectively separating the intrinsic background from the R-Cu peak emission. 
(b) PL spectra from ZnS:Cu NC (black trace) and undoped ZnS NC (grey trace) films, collected at 19 K with continuous wave, 375 nm excitation. Intensities are normalized at 430 nm. The difference spectrum (red curve) is almost identical to the time-gated spectrum from ZnS:Cu NC films under 375 nm, 500 kHz pulsed excitation (purple dashed curve).
} 
\end{figure}

Figure \ref{fig:TRES}a shows how time-resolved emission spectroscopy can be used to isolate R-Cu PL from the intrinsic background PL, since most of the intrinsic PL occurs within 250 ns of excitation while the R-Cu PL is longer lived. 
The room-temperature PL decay of ZnS:Cu NCs, excited with a pulsed excitation source at $\lambda_\mathrm{ex}$= 375 nm and monitored at $\lambda_\mathrm{em}$= 670 nm, is tri-exponential with decay time constants ($\tau_i$) of $\tau_1$=1.85$\mu$s, $\tau_2$=8.72$\mu$s, and $\tau_3$=26.47$\mu$s. 
With 95\% confidence, we find that these $\tau_i$ are consistent among all three Cu-doped samples (SI Section 6). 
At 19 K, time-resolved emission spectroscopy separates peaks I and II as well as R-Cu from the intrinsic background PL. 
Figure \ref{fig:TRES}b shows that time-gating the PL from the ZnS:Cu NC samples yields an almost identical spectral shape to that of calculating the difference between the normalized, CW PL spectra from the doped NCs and the undoped NCs. 
The CW spectra in Figure \ref{fig:TRES}b are normalized such that the PL intensities collected at 430 nm (the shortest emission wavelength in the measurement) are the same, as PL at this wavelength is expected to arise predominantly from intrinsic defects. 
The observation that nearly identical spectra are obtained, either by time-gating the doped spectrum or by subtracting the undoped CW spectrum, strongly supports that peaks I and II arise from Cu doping, along with the R-Cu peak, and these peaks coexist with the intrinsic PL in doped samples for sub-bandgap excitation.

\subsection{First-principles calculations}

\begin{figure}
\centering
\includegraphics[scale=0.85]{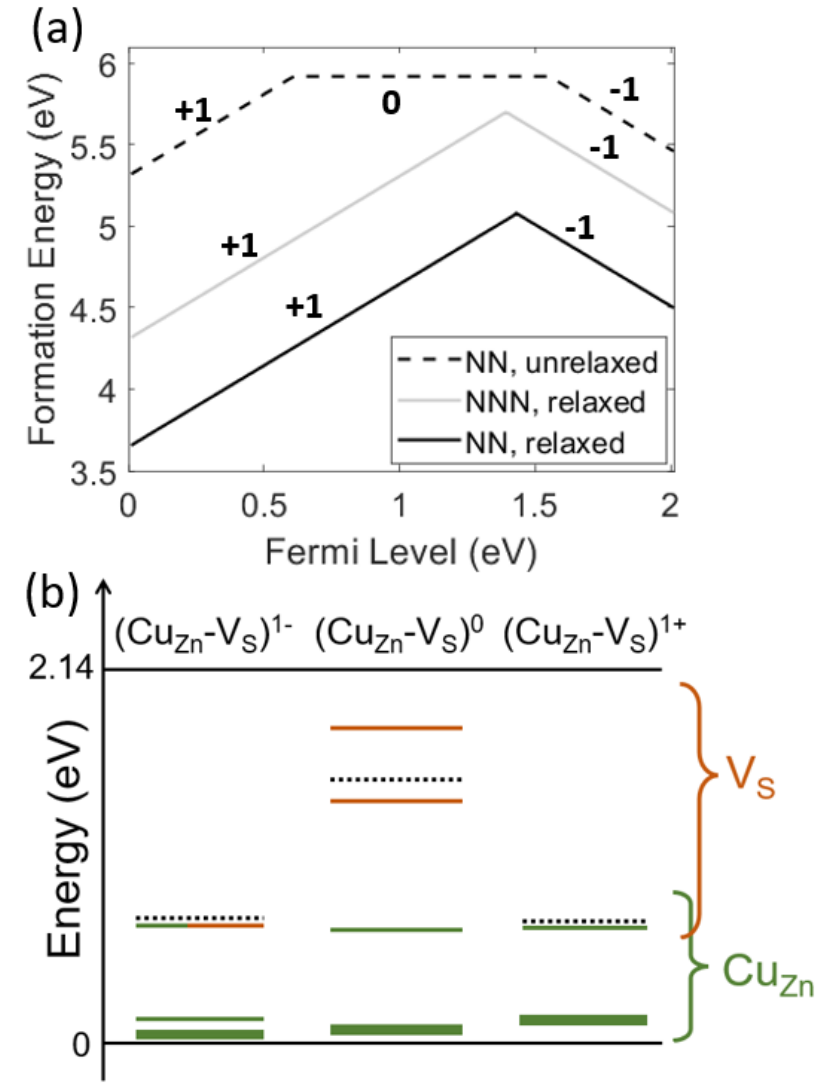}
\centering
\caption{\label{fig:DFT}
\textbf{First-principles calculations} (a) Formation energies for nearest- and next-nearest-neighbor (NN and NNN, respectively) associations of Cu\textsubscript{Zn} and V\textsubscript{S} in ZnS, as a function of the Fermi level (solid curves).
Charge states with respect to the ZnS lattice are indicated as -1, 0, and +1 for the negatively charged, neutral, and positively charged complex, respectively.
The dashed curve shows the formation energy for the NN complex in an unrelaxed ZnS lattice. All formation energy calculations are performed under zinc-rich sulfur-poor thermodynamical stability conditions. 
(b) Defect levels at $k=0$ for three different charge states of the NN Cu\textsubscript{Zn}-V\textsubscript{S} complex. 
Orange lines indicate V\textsubscript{S}-derived states, and green lines indicate Cu\textsubscript{Zn}-derived states. 
Solid lines indicate the valence band maximum (0 eV) and conduction band minimum (2.14 eV). 
Dotted lines indicate the Fermi level.
} 
\end{figure}

R-Cu PL has been proposed to arise from a nearest-neighbor (NN) complex of Cu\textsubscript{Zn} and V\textsubscript{S} defects, rather than more distant associations\cite{Shionoya1966NatureMeasurements}. 
To confirm the thermodynamic stability of the NN Cu\textsubscript{Zn}-V\textsubscript{S} complex, we use density functional theory (DFT) to calculate the formation energies, defect levels, and projected density of states (PDOS) for ground-state configurations of the complex in several charge states, as well as for the next-nearest-neighbor (NNN)  complex. 
The results of these calculations are shown in Figure \ref{fig:DFT}. 
We find that the formation energy of the NN Cu\textsubscript{Zn}-V\textsubscript{S} complex is lower than that of the NNN complex. 
The formation energy calculations in Figure \ref{fig:DFT}a indicate that the two stable charge states are either negative ($-$1) or positive ($+$1), depending on the Fermi level, with the neutral (0) configuration always lying higher in energy.
This is in contrast to the calculation for NN Cu\textsubscript{Zn}-V\textsubscript{S} in an unrelaxed ZnS lattice, which significantly increases the formation energy of all charge states, but particularly the negative and neutral configurations. 

Figure \ref{fig:DFT}b shows the defect levels and their projections at $k=0$ for each charge state of the NN complex.
These calculations qualitatively agree with the relative arrangement of levels in Figure \ref{fig:LT PL/PLE}b, with orange lines indicating the positions of two, higher-energy states derived from Zn dangling bonds surrounding the V\textsubscript{S} site, and green lines indicating ten, lower-energy states derived from the Cu\textsubscript{Zn} $d$-shell. 
The total density of states for pure ZnS and for ZnS containing a neutral Cu\textsubscript{Zn}-V\textsubscript{S} complex are included in SI Section 7.
In the negatively-charged complex, all twelve states are occupied, and the V\textsubscript{S}-derived states are strongly mixed with the Cu\textsubscript{Zn}-derived states. 
In the neutral complex, the V\textsubscript{S}-derived states are only partially filled and are no longer mixed with the Cu\textsubscript{Zn}-derived states. 
In the positively-charged complex, only the Cu\textsubscript{Zn}-derived states are occupied and the V\textsubscript{S}-derived states are no longer easily isolated; likely because they have been pushed far into the conduction band; however, this may be an artifact of well-known DFT bandgap errors (the estimated bandgap in this calculation is 2.14 eV, compared to the expected value around 3.6 eV), and the V\textsubscript{S} states may still exist within the bandgap.

For the R-Cu transition depicted in Figure \ref{fig:LT PL/PLE}b to occur, there must be a hole in the higher-energy Cu\textsubscript{Zn} states. 
This hole is likely created by photo-ionization of a Cu\textsubscript{Zn} electron into the conduction band based on the large Stokes shift we observe between peak $\lambda_\mathrm{em}$ and peak $\lambda_\mathrm{ex}$ for R-Cu PL. 
It has also been proposed that this Stokes shift is a result of lattice relaxation around the excited complex when a Cu\textsubscript{Zn} electron is transferred directly to a V\textsubscript{S} state\cite{Koda1964, Shionoya1966NatureMeasurements}.
In this case, the excited V\textsubscript{S} level lies above the conduction band minimum immediately after excitation, and may therefore release an electron to the conduction band before being lowered into the bandgap following lattice relaxation.  
If the excited complex resulting from either of the above processes contains an electron in a V\textsubscript{s} state, R-Cu emission can subsequently occur. 
Otherwise, an electron must be re-captured by the complex into a V\textsubscript{s} state for R-Cu emission to occur, leading to a longer emission lifetime. 
Based on this observation, the electron occupations of the defect levels in Figure \ref{fig:DFT}b indicate how the charge state prior to excitation determines the possible emission pathways, which define the emission lifetime and the energy of the R-Cu PL.

\subsection{R-Cu Emission Dynamics}

\begin{figure*}[ht]
\centering
\includegraphics[width=15.5cm]{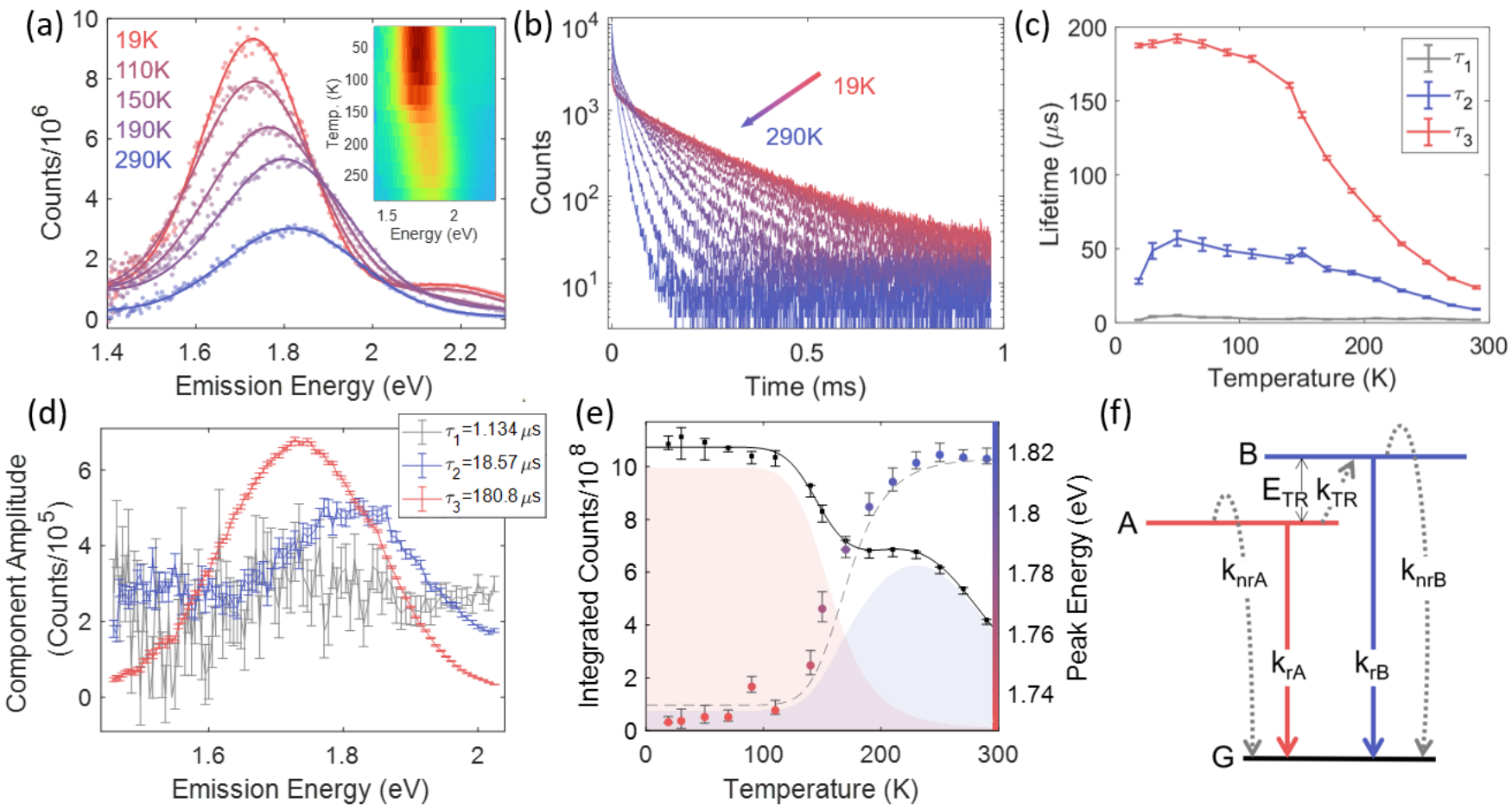}
\centering
\caption{\label{fig:Dynamics}
\textbf{R-Cu Emission Dynamics} (a) PL spectra measured at temperatures ranging from 19 K-290 K (data points for five representative temperatures are shown, with all data plotted in the inset) and Gaussian fits (solid traces).
(b) Time-dependent PL emission following pulsed excitation at $\lambda_\mathrm{ex}$=375 nm, measured at the peak PL wavelength for temperatures from 19 K to 290 K.
(c) PL decay lifetimes extracted from a tri-exponential fit to the time-dependent PL curves in (d) at each measurement temperature.
(d) Amplitudes of each tri-exponential decay component at a single temperature (19 K) as a function of emission energy.
(e) Integrated emission intensity (black points) and peak energy (colored points) as a function of temperature, extracted from the Gaussian fits of (a). Error bars represent 68\% confidence intervals from fit results.
The solid black curve is a fit to the intensity data using the model described in the text.
Red and blue shaded regions represent the relative temperature-dependent intensities $I_A(T)$ and $I_B(T)$ from the best-fit model, and the dashed curve is a sum of the two emission energies resolved in (d), weighted by their corresponding best-fit emission intensities. 
(f) Energy level diagram showing two manifolds of states inside the ZnS bandgap with coupled relaxation processes, where radiative recombination from A to G results in 1.73 eV emission and radiative recombination from B to G results in 1.82 eV emission.
} 
\end{figure*}

Figure \ref{fig:Dynamics} shows how the spectral and temporal characteristics of the R-Cu PL as a function of temperature provide detailed insight regarding the emission mechanisms and the states involved.
At temperatures from 19 K to 290 K, we measure the PL emission spectrum to find the peak $\lambda_\mathrm{em}$, and we then measure the corresponding PL decay curve for that $\lambda_\mathrm{em}$. 
The PL spectra at each temperature are converted to energy units (see Methods) and fit using Gaussian functions to extract the peak energies ($E_\mathrm{em}$) and integrated intensities. 
The emission spectra at 19 K, 110 K, 150 K, 190 K, and 290 K are plotted as examples in Figure \ref{fig:Dynamics}a along with the corresponding fit results. 
The spectral data for all measurement temperatures are shown in the pseudocolor plot of the inset, and fitted spectra for all measurement temperatures not included in Figure \ref{fig:Dynamics}a are shown in SI Section 8. 
For the PL decay measurements at each temperature (Figure \ref{fig:Dynamics}b), we find that a tri-exponential decay model most effectively describes the data compared to fitting with one, two, or four exponential terms or a stretched exponential decay function. 
The best-fit lifetime components, $\tau_i$ for $i=1$,2,3 at every temperature are plotted in Figure \ref{fig:Dynamics}c, showing three, well-separated decay lifetimes. 

At the lowest measurement temperature of 19 K (Figure \ref{fig:Dynamics}d), we acquire PL decay curves across the R-Cu emission band with 2.5 nm resolution, and we fit the data using the tri-exponential model with fixed lifetimes based on the fit results from Figure \ref{fig:Dynamics}c.
Figure \ref{fig:Dynamics}d shows the PL amplitudes corresponding to the decay components $\tau_1$, $\tau_2$, and $\tau_3$ as a function of $\lambda_\mathrm{em}$.
The fast component $\tau_1$ likely reflects the tail of one or more peaks outside the R-Cu emission band, with little spectral dependence.
However, separating the slow ($\tau_3$) and fast ($\tau_2$) PL contributions this way reveals the presence of two distinct peaks at 1.73 eV and 1.82 eV. 
The observation of energetically distinct PL peaks with different lifetimes is consistent with the co-existence of two separate radiative transitions.

Figure \ref{fig:Dynamics}e shows the integrated PL intensity over the R-Cu band and best-fit $E_\mathrm{em}$ at every temperature, extracted from Gaussian fits to the PL data in Figure \ref{fig:Dynamics}a. 
As noted previously, $E_\mathrm{em}$ blueshifts as the temperature increases, and Figure \ref{fig:Dynamics}e illustrates that the shift occurs non-linearly, with a marked inflection between 100 K and 200 K and saturation at both higher and lower temperatures.
Meanwhile, the R-Cu emission intensity decreases with increasing temperature from 19 K to 190 K, then plateaus between 190 K and 210 K, before decreasing again at higher temperatures. 
The initial decrease in intensity is consistent with quenching through thermally-activated non-radiative recombination pathways and is typical for defect emission. 
The plateau in the intensity between 150 K and 270 K is anomalous and can be explained by a temporary increase in PL intensity from a subset of radiative states, referred to as negative thermal quenching (NTQ). NTQ is occasionally observed in defect emission; for example, in the case of the 2.65 eV PL (referred to as the YS1 peak) from ZnS:I\cite{Yokogawa1983}, and it has generally been explained by thermally-activated carrier transfer from lower- to higher- energy emissive defect states. 
ZnS:Cu NCs have been synthesized in water at room temperature with the same Cu(CH\textsubscript{3}COO)\textsubscript{2} precursor\cite{Bol2002} and then subsequently annealed at 450 $^{\circ}$C to intensify R-Cu PL, but the resulting red peak (600 nm at room temperature) is not resolvable from other emission peaks when the temperature is less than 220 K, making it impossible to observe a similar NTQ or blueshift. 

In Figure \ref{fig:Dynamics}f, we propose an empirical model to capture both the temperature-dependent blueshift and the NTQ of R-Cu emission.
Motivated by the time-resolved observations of Figure \ref{fig:Dynamics}d, we include two radiative recombination transitions with emission energies at  1.73 eV (A$\rightarrow$G) and 1.82 eV (B$\rightarrow$G), corresponding to two distinct excited-state configurations.
These radiative transitions compete with thermally-activated, non-radiative transitions that generally tend to quench the emission at elevated temperatures.
However, as carriers are thermally excited from state A to state B at temperatures with thermal energy corresponding to the energy offset $E_\mathrm{TR}$, the faster B$\rightarrow$G transition increasingly becomes the dominant radiative recombination pathway, resulting in blueshifted PL and temporary NTQ. 
This mechanism is consistent with our observation that inflection points in the PL intensity align with the onset and saturation of the blueshift in $E_\mathrm{em}$.

To quantify this model, we derive the following analytical expressions for the temperature-dependent PL intensities, $I(A)$ and $I(B)$, from the radiative transitions occurring from excited states A and B, respectively:
\begin{gather}
    I_A(T) = I_A(0)\frac{k_\mathrm{rA}}{k_\mathrm{rA} + k_\mathrm{nrA} + k_\mathrm{TR} },  
    \label{eq:ita} \\
I_B(T) = I_B(0)\frac{k_\mathrm{rB}}{ k_\mathrm{rB} + k_\mathrm{nrB} } + I_A(0) \frac{ k_\mathrm{TR} k_\mathrm{rB} }{ k_\mathrm{rB} + k_\mathrm{nrB} }. 
\label{eq:itb}
\end{gather}
Here, $k_\mathrm{rA}$ and $k_\mathrm{rB}$ are the radiative recombination rates shown in Figure \ref{fig:Dynamics}f (solid lines), which are independent of temperature.
The terms $k_\mathrm{nrA}$ and $k_\mathrm{nrB}$ are rates for non-radiative relaxation, and $k_\mathrm{TR}$ is the rate for non-radiative transfer between states A and B (dashed lines in Figure \ref{fig:Dynamics}f).
These non-radiative rates are temperature-dependent with the form $k_j = \Gamma_j \exp(-E_j/k_\mathrm{B} T)$, where $\Gamma_j$ is a proportionality constant, $E_j$ is the activation energy of the transition, and $k_\mathrm{B}$ is Boltzmann's constant.
See SI Section 9 for a derivation of these expressions.

The sum of Equations (\ref{eq:ita}) and (\ref{eq:itb}) gives the total PL intensity as a function of temperature. This expression is used as a model to fit the temperature-dependent PL intensity data in Figure \ref{fig:Dynamics}e, from which we extract best-fit values for the energies $E_\mathrm{nrA}$, $E_\mathrm{nrB}$, and $E_\mathrm{TR}$.
The best-fit value of $E_\mathrm{TR}$ is 153$\pm$22 meV. 
Details of the fitting procedure along with best-fit results for other parameters are included in SI Section 9.
It is worth mentioning that our expression for the temperature-dependent PL intensity resembles an Arrhenius equation with a NTQ term, as in the case of ZnS:I\cite{Shibata1999NegativeSolids}, and we have concluded that it is not possible to obtain an intensity curve with the measured temperature dependence by considering an Arrhenius equation with only positive thermal quenching terms. 
To confirm the validity of this model, the dashed curve in Figure \ref{fig:Dynamics}e represents a weighted sum of the 1.73 eV and 1.82 eV PL contributions, with weights determined by the best-fit $I_A(T)$ and $I_B(T)$ values (the intensities are represented by shaded regions in Figure \ref{fig:Dynamics}e).
We recover a temperature-dependent emission energy that closely tracks the measured $\sim$90 meV blueshift of $E_\mathrm{em}$ between 19 K and 290 K.

The states A and B in our empirical model might be associated with different charge states or spatial configurations of Cu\textsubscript{Zn} and V\textsubscript{S}. 
Based on the defect level calculations in Figure \ref{fig:DFT}b, the energy levels associated with both Cu\textsubscript{Zn} and V\textsubscript{S} shift in different charge states, as do the degree of orbital hybridization between Cu\textsubscript{Zn} and V\textsubscript{S} states.
The emission energies and lifetimes associated with different charge states should therefore be different.
(Note, however, that this remains a qualitative observation since the ground-state PDOS calculations do not fully capture the energies of excited-state configurations, which would be required to calculate emission energies.)
We also consider the possibility that defects outside of Cu\textsubscript{Zn}-V\textsubscript{S} complexes are responsible for activating R-Cu at low or high temperatures.
Possible candidates for defects creating donor levels in ZnS include Zn\textsubscript{i} or Cu\textsubscript{i} interstitial defects, as well as lone V\textsubscript{S} defects.
Cu\textsubscript{i} is a shallow donor in ZnS\cite{Hoang2019} and is therefore unlikely to play a role as the donor level in R-Cu emission at any temperature. 
At low temperatures, it is possible that the primary donors in the R-Cu emission mechanism are lone V\textsubscript{S} defects in the NC core or at the surface, because their more distant interaction with Cu\textsubscript{Zn} relative to V\textsubscript{S} in a Cu\textsubscript{Zn}-V\textsubscript{S} complex would be consistent with longer-lived and lower-energy radiative recombination.
The inverse situation, in which lone Cu\textsubscript{Zn} defects are primary acceptors at low temperatures, would be similar in a model considering holes instead of electrons; however, this situation is less likely given the high concentration of V\textsubscript{S} defects expected to exist at the NC surface compared to the Cu\textsubscript{Zn} defects in these samples.
Ultimately, carrier transfer between two manifolds of states may be advantageous for potential defect qubit architectures if it is found to be spin-dependent, or mitigated if it is found to be detrimental\cite{Hopper2018}.

We also considered other potential explanations for the R-Cu blueshift and NTQ.
Previous reports of blueshifted R-Cu emission upon increasing temperature from bulk ZnS:Cu were attributed to changing occupation in the vibrational levels of a highly-localized center\cite{Shionoya1964NatureCrystals}. 
However, that explanation is not consistent with the saturation in the blueshift at high temperatures, which we clearly observe and which also appears to occur in their measurement around 200 K. 
Characteristic defect PL in bulk and nanocrystalline ZnS:Mn also exhibits a blueshift upon increasing temperature with magnitudes between 25 meV and 80 meV, which has been attributed to crystal field variations due to lattice expansion \cite{Chen2002TemperatureNanoparticles, Park2007}. 
The crystal-field explanation would also not predict saturating behavior, and accordingly the temperature-dependent blueshift of ZnS:Mn defect PL does not saturate, in contrast to the R-Cu observations. 
As an additional indication that crystal field effects cannot sufficiently explain the measured R-Cu blueshift, we calculate only a 16 $\pm$ 0.01 meV increase in the energy separation between Cu\textsubscript{Zn} and V\textsubscript{S} levels of the neutral complex upon performing DFT computations with different ZnS lattice constants, corresponding to 0 K and 300 K based on the ZnS thermal expansion coefficient\cite{Wang2006}. 
It is worth noting that in nanocrystalline ZnS:Mn, NTQ has also been reported between 50 K and 300 K, with positive thermal quenching resuming above 300 K\cite{Park2007}. 
The authors attributed the NTQ to the thermal depopulation of localized trap states created by lattice defects, organic impurities, and surface defects which are all expected to be more prevalent in NCs compared to bulk materials. 
None of the above alternative models for the R-Cu blueshift and NTQ can explain the clear presence of two distinct emission peaks with different radiative lifetimes at low temperature, as we observe in Figure \ref{fig:Dynamics}d, nor do they capture the correspondence in Figure \ref{fig:Dynamics}e between the NTQ regime and the blueshift occurring at approximately the same temperature.

\section*{Conclusion}
We present a synthetic method for obtaining colloidal ZnS:Cu NCs that emit primarily R-Cu, with a tailorable intensity depending on the Cu concentration. 
Using time- and temperature- resolved measurements and first-principles calculations, we find the sub-bandgap PL is consistent with radiative transitions from two, coupled manifolds of states involving Cu\textsubscript{Zn}-V\textsubscript{S} complexes. 
In the future, experimental methods unique to colloidal NC platforms will clarify the importance of defect type, location, concentration, and charge state on R-Cu PL. 
For example, post-synthesis NC modification (\textit{e.g}., surface treatments that passivate traps and for remote doping, cation exchange, or sulfidation) and the growth of core-shell heterostructures will controllably alter the environments and compositions of defects as well as the Fermi level of NCs.  
Spectroelectrochemical measurements of colloidal NC dispersions can also reveal the relationship between defect PL and the NC Fermi level.
These studies, combined with ESR and ODMR measurements that probe spin transitions, will yield valuable information about the R-Cu electronic structure and spin-dependent optical properties for potential development of R-Cu centers as defect qubits.

Colloidal NC hosts will also facilitate the isolation and study of individual R-Cu color centers, compared to bulk hosts which have typically been used for the development of color centers as defect qubits. 
The deposition of sparse dispersions of colloidal NCs reduces the bulk purity requirement for single-quantum-emitter measurements by thousands of times\cite{Kagan2020ColloidalScience,Shulevitz2022}.
Furthermore, established methods for colloidal NC luminescence enhancement by integration with resonant photonic cavities or plasmonic nanostructures can improve the efficiency of quantum emitter measurements by reducing PL lifetimes through Purcell enhancement\cite{Saboktakin2013, Raha2020}.

Our investigation of R-Cu color centers also motivates studying other transition-metal-vacancy complexes in ZnS as potential defect qubits; for example, transition metals with fewer $d$-shell electrons than Cu, when placed in a similar defect and charge configuration, could produce a higher ground-state spin and a greater number of internal radiative transitions. 
Such theoretically interesting materials systems are readily available for experimentation through colloidal NC synthesis methods, which are fast and accessible compared to methods that exist for bulk crystals and can be atomically precise\cite{Muckel2016}.  
All of the above opportunities will facilitate the development of color centers in ZnS as quantum defects while generally motivating colloidal NCs as a hosts for quantum defect development and engineering.

\section{Methods}

\subsubsection{Synthesis of Colloidal ZnS:Cu NCs}

A 10 mL solution of Cu(CH\textsubscript{3}COO)\textsubscript{2}$\cdot$H\textsubscript{2}O in DI water is prepared with the appropriate molar concentration of Cu, i.e.,  0.05\%, 0.075\%, or 0.1\% of the molar concentration of Zn in the reaction. 0.1 mL of this solution is then added to a 50 mL three-necked flask containing 20 mmol OM. 
The mixture is degassed for 45 min at 120 $^{\circ}$C before the injection of 20 mmol OA and 0.2 mmol Zn(Ddtc)\textsubscript{2}, followed by 45 min additional degassing. 
The vessel is then heated to 300 $^{\circ}$C under a nitrogen atmosphere and maintained at 300 $^{\circ}$C for 45 min. 
It is then left to cool to 60 $^{\circ}$C. 
The cooled contents are mixed with excess ethanol, and NCs are collected \emph{via} centrifugation, washed in ethanol, and re-dispersed in hexanes to a concentration of 10 mg/mL. 

\subsubsection{Tools and Instrumentation}

ICP-OES measurements are collected using a SPECTRO GENESIS ICP-OES spectrometer. 
To collect TEM images, 2 mg/mL NC dispersions in hexanes are drop-cast onto carbon-coated copper grids and imaged using a JEOL-1400 TEM. 
TEM images are analyzed using Fiji\cite{Schindelin2019}.
Absorption spectra are measured using an Agilent Cary 5000 spectrometer. 
PL and PLE spectra are measured using an Edinburgh Instruments FLS1000 spectrometer with a PMT-980 photodetector. 
For continuous PL and PLE measurements, the excitation source is a 450W Xe lamp. 
For time-resolved measurements, the excitation source is a 375 nm Picoquant LDH-series laser diode. 
For temperature-controlled measurements, samples are placed in an evacuated Advanced Research Systems DE-202 cryostat.
The illustration of a spherical ZnS NC in the Table of Contents graphic was generated using NanoCrystal\cite{Chatzigoulas2018}.

\subsubsection{Analysis of PL Emission Spectra}

PL spectra are measured as a distribution function of wavelength and converted to energy units prior to Gaussian fitting to extract the positions and widths of individual peaks. 
To properly account for the nonlinear relationship between wavelength and energy, we scale the spectra using the appropriate Jacobian transformation\cite{Mooney2013}:
\begin{equation}
    f(E)=f(\lambda)\frac{hc}{E^2}
\end{equation}
The broad spectral range and large peak widths in our measurements make this scaling factor critical in our analysis. 
We find that simply converting the peak wavelengths in the as-measured spectra to energy units would result in the extraction of dramatically incorrect peak energies. 
This can be seen in the results of Table \ref{table:peaks}, which take the scaling factor into account prior to peak extraction on an energy scale.

\subsubsection{Computational Details}

The electronic structures of the bulk ZnS, and defect centers such as Cu\textsubscript{Zn} and Cu\textsubscript{Zn}-V\textsubscript{S} complexes in ZnS are studied using density functional theory (DFT) with the Vienna Simulation Package\cite{vasp1} (VASP). VASP employs the Perdew-Burke-Ernzerhof (PBE) functional for the exchange and correlation within the augmented plane wave (PAW) scheme \cite{vasp2,KressePRB99}. 
For the supercell, we use two different sizes: one with 64 atoms and the other with 212 atoms. 
Both calculations yield the same results for formation energies, electronic band structure, and total and projected DOS. 
We use a total energy cut-off of 300 eV, and 6x6x6 and 12x12x12 Monkhorst-Pack k-point meshes for the density of states calculations in the larger and smaller supercells, respectively.

The formation energies of differently charged configurations are calculated from the well-known defect formula, \cite{Zhang1991}
\begin{align}
    E_{f}^q(\epsilon_F)=E^q_{\text{tot}}-E^{\text{bulk}}_{\text{tot}}&+E_{\text{corr}}+ \sum_i n_i \mu_i\\&
   \nonumber +q(E_{\text{VBM}}+\epsilon_F+\Delta_{q/b} )
\end{align}
where the first two terms are the total energies of the bulk and defected supercell, and the correction term $E_{\text{corr}}$ (first order Makov-Payne correction) originates from the interaction between periodic charged supercells. 
The chemical potentials $\mu_i$ correspond to adding Cu or removing Zn and S, $\epsilon_F$ is the Fermi level, $E_{\text{VBM}}$ is the valence band maximum. 
The final term $\Delta_{q/b}$ is the potential alignment between the valence band edges for the bulk and neutral or charged supercells. 

Hybrid DFT calculations are also completed for the 64 atom supercell using the B3LYP hybrid functional. 
Charge transition levels of the formation energies are not affected by the hybrid calculation, but the conduction band minimum is pushed from about 2.14 eV to 3.55 eV, yielding a band gap energy closer to that which is observed experimentally (SI Section 10).

\section{Financial interest statement}
The authors declare no competing financial interest.

\begin{acknowledgements}
This work was supported by the National Science Foundation under Awards DMR-2019444 (S.Y., C.B.M., L.C.B., and C.R.K., for synthesis, measurements, and analysis), and DMREF awards DMR-1922278 (L.C.B) and DMR-1921877 (C.S. and M. E. F.) for theory, first-principles calculations, and data analysis.
S.M.T. acknowledges support from the National Science Foundation Graduate Research Fellowship under Grant No. DGE-1845298.

\end{acknowledgements}


\bibliography{Bib}
\end{document}


\pagenumbering{arabic}

\section{1. Nonnegative Matrix Factorization of Room-Temperature Emission Spectra}
We use the built-in nnmf function in MATLAB R2021b to decompose room-temperature PL emission spectra from differently-doped ZnS:Cu NCs. The input matrix for factorization contains data from all materials and the rank of factors is 2. The relative weights of the decomposition components in each spectrum are used to plot the relative strength of the red spectral component in Figure 1b of the main text.

	\begin{figure}[!h]
      \begin{center}
          \includegraphics[width=13cm]{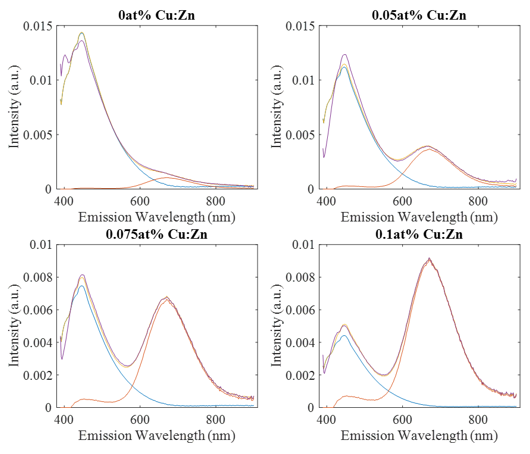}
          \caption{NNMF decomposition of room-temperature PL emission spectra under 375 nm excitation from ZnS and ZnS:Cu NCs synthesized with 0 mol\%, 0.05 mol\%, 0.075 mol\%, and 0.1 mol\% Cu:Zn. The two spectral components in the factorization are plotted in blue and orange, with their sum plotted in yellow and the measured data plotted in purple.}
          \label{nnmf}
      \end{center}
    \end{figure}
    
\clearpage
\newpage

\section{2. Measurement and Calculation of NC Bandgap Energies}

\begin{table}[h!]
\begin{tabular}{cccc}
\hline
Cu:Zn mol\% & NC Size (nm) & Avg. Calculated Bandgap (eV) & Measured Bandgap (eV) \\   \hline
 0 &  7.0 $\pm$ 1.3 &  3.85  & 3.79 \\
 0.05 & 7.4 $\pm$ 1.2   &  3.84 & 3.77  \\  
 0.075 & 7.2 $\pm$ 1.1 & 3.85 & 3.79   \\ \
 0.1 & 7.5 $\pm$ 1.2 & 3.83  & 3.79   \\    
 \hline
\end{tabular}
\caption {\label{tab:sizebg} Size distribution and bandgap energies for Cu-doped ZnS NCs synthesized with four different Cu:Zn molar ratios.} 
\end{table}

 
\begin{figure}[!h]
\begin{center}
  \includegraphics[width=10cm]{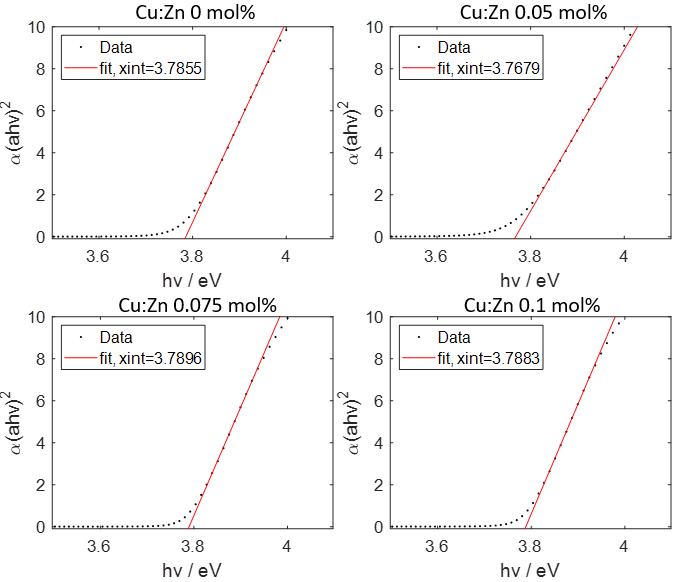}
  \caption{Tauc plots of absorption spectra data for extraction of NC bandgap energies.}
  \label{nnmf}
\end{center}
\end{figure}

\clearpage
\newpage

The average NC radius according to TEM image analysis is used to calculate a bandgap energy for each sample using Equation \ref{eq:bg}, in which R is the NC radius, $m_e^*$ is the effective electron mass (0.25$m_0$), $m_h^*$ is the effective hole mass (0.59$m_0$), $\epsilon_r$ is the relative permittivity (8.9), and $E_g$ is the bandgap energy of bulk ZnS (3.68 eV)\cite{Uzar2011}.

 \begin{equation} \label{eq:bg}
E\textsubscript{g} + \frac{\pi^2\hbar^2}{2m_0R^2}(\frac{1}{m_e^*} + \frac{1}{m_h^*}) - \frac{1.8e^2}{4\pi\epsilon_r\epsilon_0R}
\end{equation}

	\begin{figure}[!h]
      \begin{center}
          \includegraphics[width=15cm]{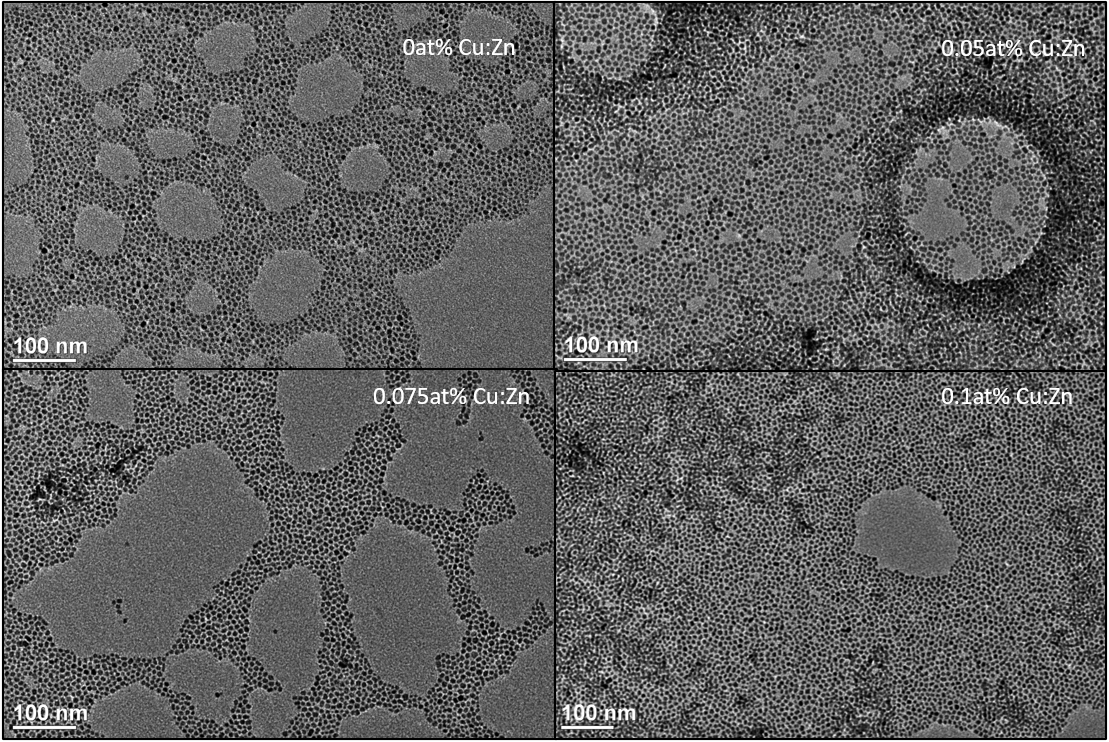}
          \caption{TEM images of ZnS NCs synthesized with 0 mol\%, 0.05 mol\%, 0.075 mol\%, and 0.1 mol\% Cu:Zn.}
          \label{tem}
      \end{center}
    \end{figure}

\clearpage
\newpage

\section{3. Measurement Conditions of R-Cu PLE Spectra}
\begin{figure}[!h]
      \begin{center}
          \includegraphics[width=14cm]{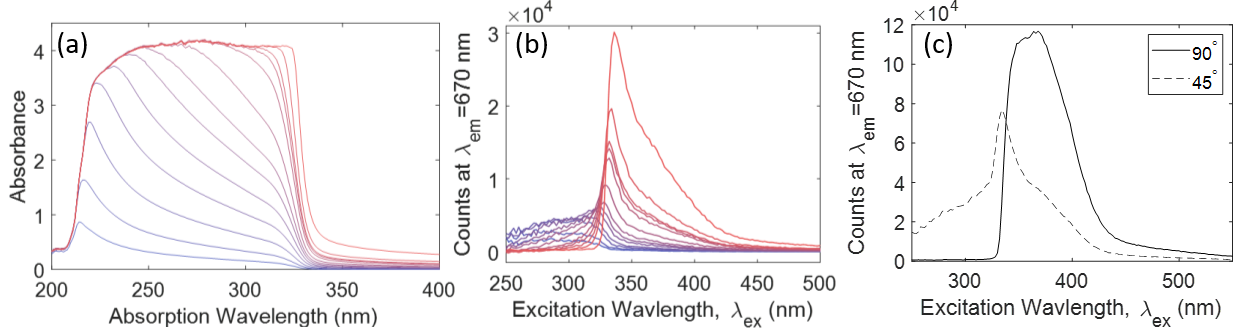}
              \caption{(a) Absorbance and (b) $\lambda_{em}$=670 nm PLE spectra measured from a dispersion containing incremental additions of NCs. The color gradient indicates the number of times NCs have been added to the sample, with blue corresponding to the lowest NC concentration and red corresponding to the highest NC concentration. (c) $\lambda_{em}$=670 nm PLE spectra measured from a 10 mg/mL dispersion of Cu:Zn=0.1mol\% ZnS:Cu NCs in hexanes in two different measurement configurations, with the cuvette rotated so that excitation is normally incident (solid curve) or incident with a 45$^\circ$ angle (dashed curve) on the face of the cuvette, such that the focal point of the measurement moves from the center to the edge of the cuvette.}
          \label{fig:concentration}
      \end{center}
    \end{figure}

We show in Figure \ref{fig:concentration}a-b that as ZnS:Cu NC samples are prepared with increasingly high concentrations, corresponding to increasingly high absorbance at above-bandgap wavelengths, the measured R-Cu PLE spectrum indicates artificially diminished above-bandgap PLE. Simultaneously, the overall measured R-Cu signal becomes intensified. As further evidence of this mechanism, we show in Figure \ref{fig:concentration}c that for a single NC concentration, changing the measurement path length through the NC dispersion by rotating the cuvette relative to the optical axes of the excitation and collection paths dramatically influences the PLE spectrum.  

\clearpage
\newpage

\section{4. Effects of Surface Treatments on Red PL}
\begin{figure}[!h]
      \begin{center}
          \includegraphics[width=15cm]{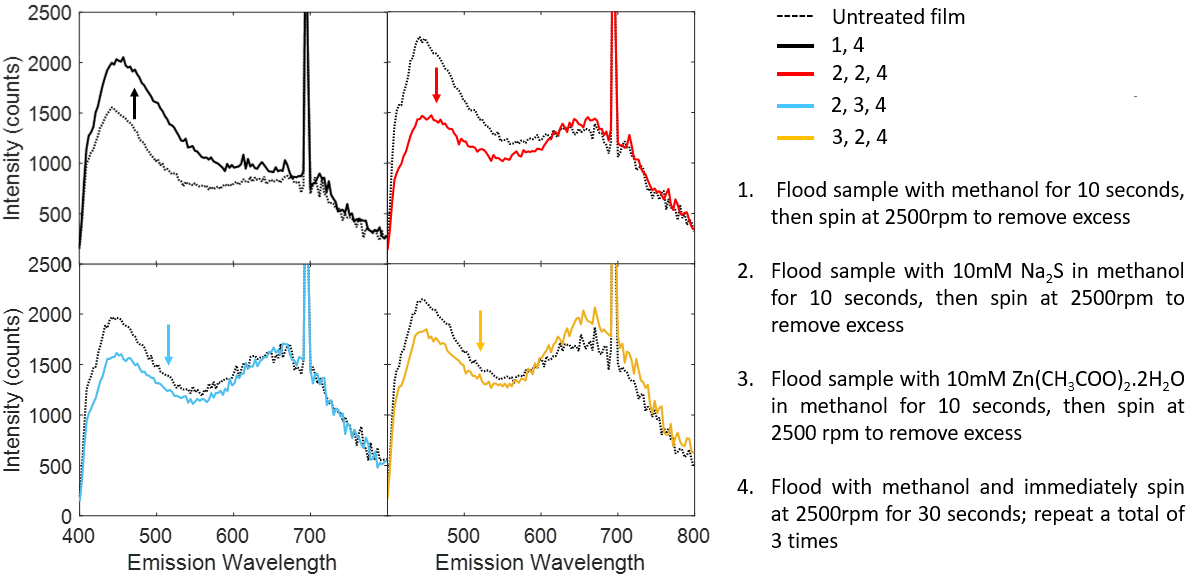}
              \caption{Room-temperature PL spectra under 375 nm excitation from ZnS:Cu NC films before and after treatments in methanol solutions containing either Na\textsubscript{2}S or Zn(CH\textsubscript{3}COO)\textsubscript{2}.H\textsubscript{2}O.}
          \label{fig:surfacetreatment}
      \end{center}
    \end{figure}

To measure the effects of altering the presence or environment of Cu cations if they are on the surface, we deposit NC solids on MPTS-treated Si wafer substrates and treat them according to the description in Figure \ref{fig:surfacetreatment}. The treatments we use involve soaking NC films in  methanol and methanolic Na\textsubscript{2}S and Zn(CO\textsubscript{2}CH\textsubscript{3})\textsubscript{2}$\cdot$2H\textsubscript{2}O solutions. Methanol is known to remove organic ligands and to strip surface cations\cite{Goodwin2014}, and methanolic Na\textsubscript{2}S and Zn(CO\textsubscript{2}CH\textsubscript{3})\textsubscript{2}$\cdot$2H\textsubscript{2}O solutions are known to enrich the NC surface in S\textsuperscript{2-} or Zn\textsuperscript{2+}, respectively\cite{Oh2014, Kim2013}

\clearpage
\newpage

\section{5. 2D Plots of PL/PLE Data}
Spectral data used to construct the PL/PLE maps in Figure 3 of the main text are shown here as 2D plots to provide additional insight.

	\begin{figure}[!h]
      \begin{center}
          \includegraphics[width=13cm]{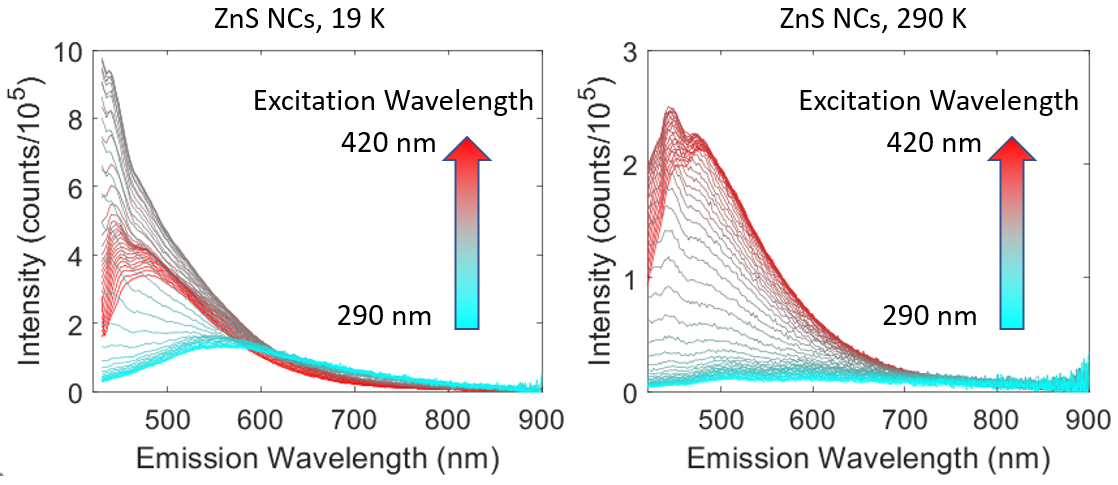}
          \caption{PL spectra from undoped ZnS NCs, measured at 19 K and 290 K under excitation wavelengths from 290 nm to 420 nm. }
          \label{plmapplot}
      \end{center}
    \end{figure}
    
	\begin{figure}[!h]
      \begin{center}
          \includegraphics[width=13cm]{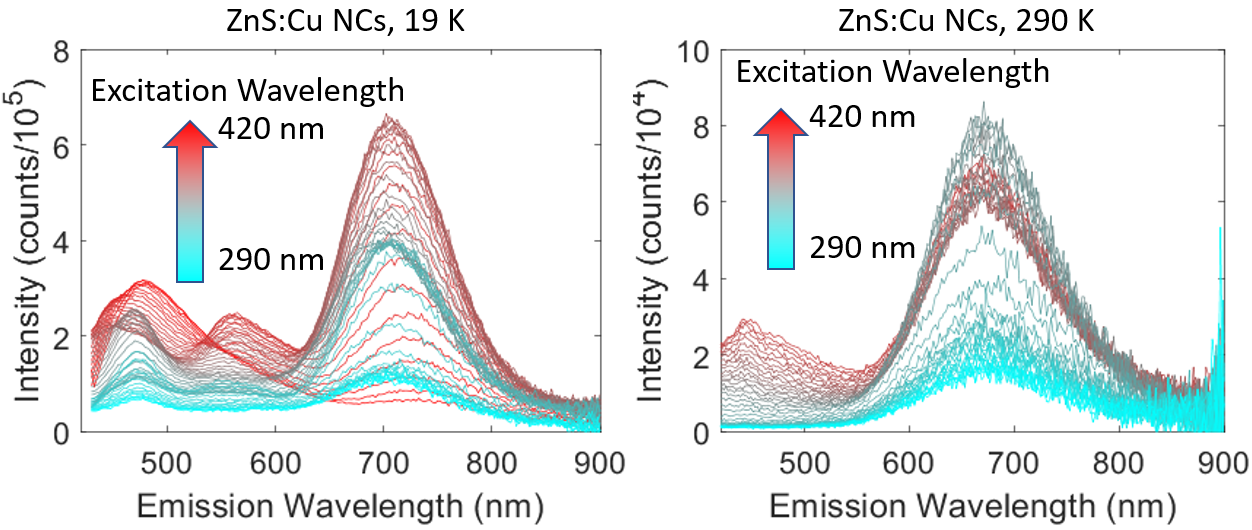}
          \caption{PL spectra from ZnS:Cu NCs, measured at 19 K and 290 K under excitation wavelengths from 290 nm to 420 nm. }
          \label{plmapplotdoped}
      \end{center}
    \end{figure}

\clearpage
\newpage

\section{6. Room-Temperature Lifetime Measurements for All Copper Concentrations}
\begin{figure}[!h]
      \begin{center}
          \includegraphics[width=15cm]{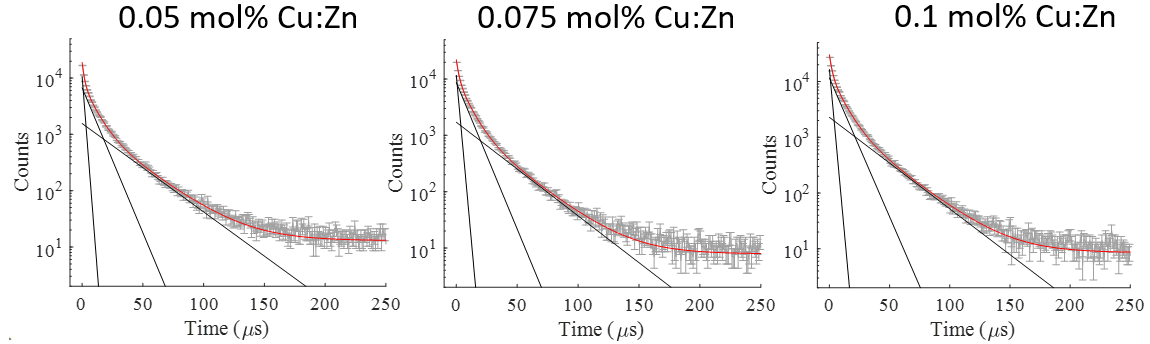}
            \caption{Room-temperature, 670 nm PL decay under 1 kHz, pulsed, 375 nm excitation (gray) with the full tri-exponential fit plotted in red and the three separate decay components plotted in black. Error bars represent the square root of the number of counts.}
          \label{rtlt}
      \end{center}
    \end{figure}

The room-temperature PL decay of the 670 nm emission peak from each NC dispersion (Figure \ref{rtlt}) was measured using a PMT-980 photomultiplier (standard for the FLS1000 Photoluminescence Spectrometer), a 5 nm monochromator collection bandwidth, and 1 kHz, 375 nm excitation. The parameters used to fit each signal to a tri-exponential decay (\ref{triexp}) are given in Table \ref{fitparams} with 95\% confidence intervals given in Table \ref{fitparamsconf}. The lifetimes $\tau_1$, $\tau_2$, and $\tau_3$ extracted from these fits are consistent across samples, indicating that changing the copper concentration in this doping range has not noticeably altered the recombination kinetics for the associated red PL emission.

\begin{equation} \label{triexp}
    I(t) = a_1(t)e^{-t/\tau_1} + a_2(t)e^{-t/\tau_2}  + a_3(t)e^{-t/\tau_3} + n
\end{equation}

\clearpage

\begin{table}[t]
\centering
\caption{Tri-exponential fit parameters for room-temperature PL decay.}
\label{fitparams}
\begin{tabular}{ |c|c|c|c| } 
\hline
Fit Parameter & 0.05 mol\% Cu:Zn & 0.075 mol\% Cu:Zn & 0.1 mol\% Cu:Zn \\
\hline
$\tau_1$ ($\mu$s) &  1.57 & 1.83  & 1.85  \\ 
$\tau_2$ ($\mu$s) & 8.43  & 8.33  & 8.72  \\ 
$\tau_3$ ($\mu$s) & 27.65  & 26.1 & 26.47 \\ \hline
$a_1$ (counts) & 1.07$\times10^4$  & 1.17$\times10^4$  & 1.64$\times10^4$ \\ 
$a_2$(counts) & 8746 & 6734  & 1.158$\times10^4$  \\ 
$a_3$ (counts) & 1561 & 1715 & 2278  \\ \hline
$n$ (counts) & 12.95  & 7.844  & 8.538 \\
\hline
\end{tabular}
\end{table}

\begin{table}[t]
\centering
\caption{95\% confidence bounds of tri-exponential fit parameters for room-temperature PL decay. }
\label{fitparamsconf}
\begin{tabular}{ |c|c|c|c| } 
\hline
Fit Parameter & 0.05 mol\% Cu:Zn & 0.075 mol\% Cu:Zn & 0.1 mol\% Cu:Zn \\
\hline
$\tau_1$ ($\mu$s) &  1.44--1.70 & 1.67--1.99 &1.72--1.97 \\ 
$\tau_2$ ($\mu$s) & 7.93--8.92& 7.85--8.82 & 8.32--9.13 \\ 
$\tau_3$ ($\mu$s) & 26.62--28.68 & 25.08--27.12& 25.62--27.33 \\ \hline
$a_1$ (counts) &1.03$\times10^4$--1.12$\times10^4$ & 1.11$\times10^4$--1.23$\times10^4$ & 1.58$\times10^4$--1.70$\times10^4$ \\ 
$a_2$(counts) & 8245--9247 & 6386--7081 & 1.108$\times10^4$--1.209$\times10^4$ \\ 
$a_3$ (counts) & 1401--1720 & 1518--1911 & 2057--2500 \\ \hline
$n$ (counts) & 12.68--13.22 & 7.621--8.066 &8.317--8.759 \\
\hline
\end{tabular}
\end{table}

\clearpage
\newpage

\section{7. Total Density of States of Pure and Defected ZnS}
	\begin{figure}[!h]
      \begin{center}
          \includegraphics[width=15cm]{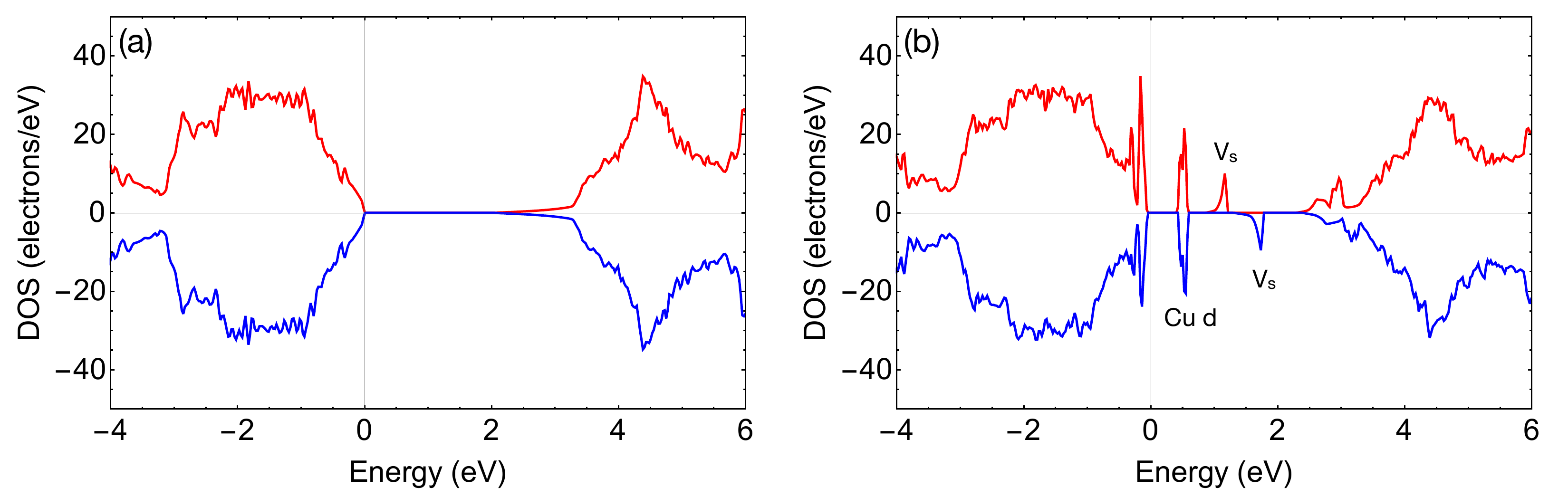}
              \caption{The total DOS of (a) pure ZnS (b) ZnS with neutral Cu\textsubscript{Zn}-V\textsubscript{S} complex. The Cu\textsubscript{Zn} $d$-levels are closer the the valence band maximum and V\textsubscript{S} states are split into two distinct energies due to the nonzero total magnetic moment in the system introduced by the Cu\textsubscript{Zn} impurity. Here zero of the energy is chosen as the top of the valence band of pure ZnS.}
          \label{table1}
      \end{center}
    \end{figure}
\clearpage
\newpage

\section{8. Fitting R-Cu Temperature-Dependent Spectra}
PL spectra are fit to two Gaussian peaks at each measurement temperature from 19 K to 290 K. Signal data are measured in counts per unit wavelength and converted to counts per unit energy for Gaussian fitting. For this conversion, the signal intensities are multiplied by a Jacobian transformation factor.\cite{Mooney2013} The fits are obtained using a weighted least squares analysis, where the weights are the uncertainty at each data point according to the assumption that measurement uncertainty is dominated by shot noise. The uncertainty at each point is therefore taken to be the Poisson variance, which is simply the number of counts, before any correction has been applied to the signal to account for the variable efficiency of the detector across wavelengths. 

\begin{figure}[!h]
	\begin{center}
          \includegraphics[width=15cm]{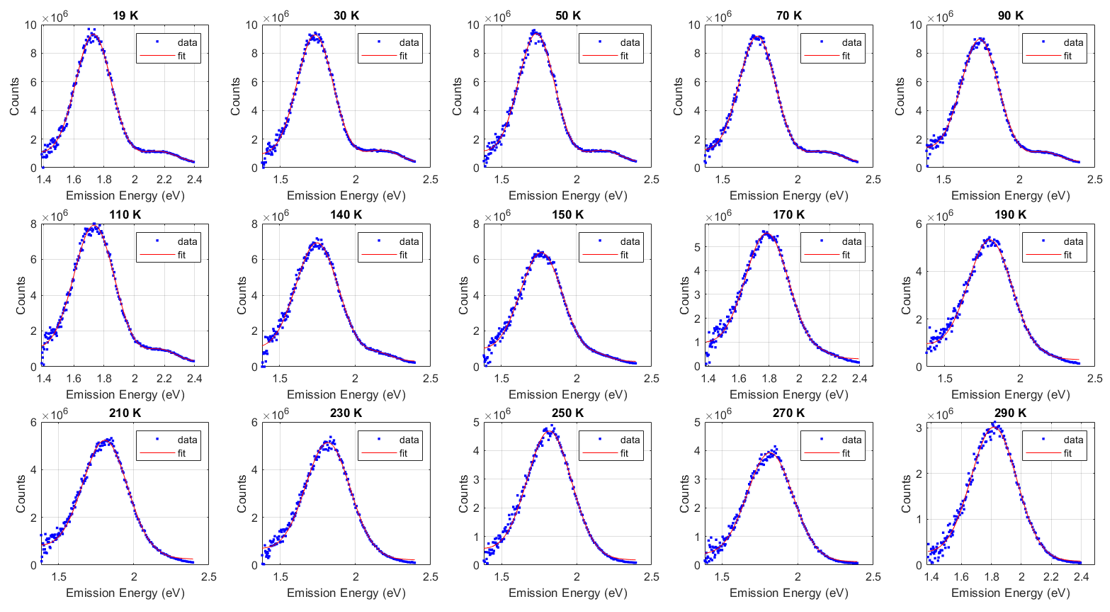}
          \caption{PL spectra measured at 19 K--290 K (blue circles) and corresponding Gaussian fits (red lines).}
          \label{rcufits}
      \end{center}
    \end{figure}

\begin{figure}[!h]
      \begin{center}
          \includegraphics[width=6cm]{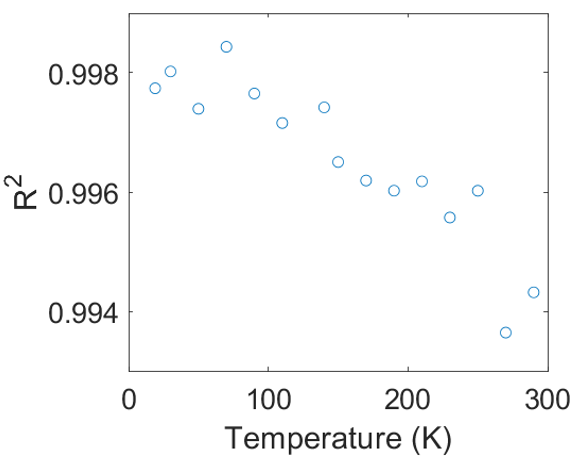}
          \caption{$R^2$ values for the Gaussian fits in Supplemental Figure \ref{rcufits} as a function of measurement temperature.}
          \label{rsq}
      \end{center}
    \end{figure}

The dominant peak between 1.73 eV and 1.82 eV at all temperatures corresponds to R-Cu emission and the higher-energy peak at approximately 2.2 eV corresponds to peak II in the main text. The fit results for all measured spectra are plotted below in Supporting Information Figure \ref{rcufits}. The $R^2$ values for each fit are given in Supporting Information Figure \ref{rsq}. 

\clearpage
\newpage

\section{9. Derivation of NTQ Equation and Best-Fit Results}

	\begin{figure}[!h]
      \begin{center}
          \includegraphics[width=5cm]{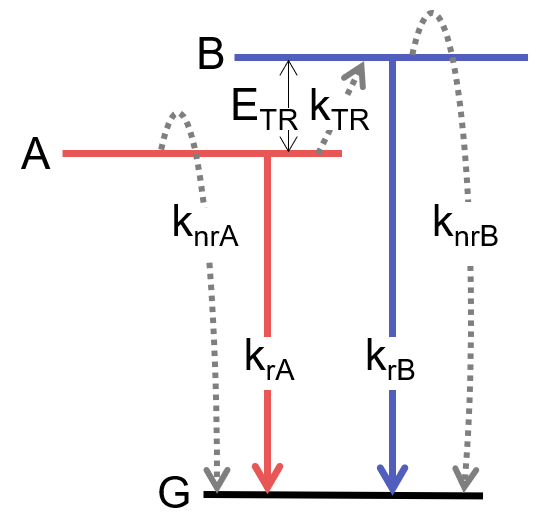}
              \caption{Model electronic structure which is proposed to produce the measured R-Cu emission dynamics. Transitions occur between a ground state, G, and two excited states, A and B. Solid arrows indicate radiative transitions, and dashed arrows indicate nonradiative transitions. Labels accompanying each arrow are transition rates. Thermal carrier transfer from A to B occurs with a rate $k_{TR}$ and activation energy $E_{TR}$ and is key in producing the measured NTQ. }
          \label{fig:derivschem}
      \end{center}
    \end{figure}

First, we  define the time derivatives of n\textsubscript{A}(T) and n\textsubscript{B}(T), which represent the electron populations of states A and B as functions of time (t) and temperature (T), using equations \ref{eq:Adef} and \ref{eq:Bdef}:

\begin{equation}
(\frac{\partial}{\partial t})n_A(t,T)= G_A(t,T) - n_A(t,T)(k_{rA} + k_{nrA}) - n_A(t,T)k_{TR}
\label{eq:Adef}
\end{equation}
\begin{equation}
(\frac{\partial}{\partial t})n_B(t,T)= G_B(t,T) - n_B(t,T)(k_{rB} + k_{nrB}) + n_A(t,T)k_{TR}
\label{eq:Bdef}
\end{equation}
\newline
where k\textsubscript{rA} and k\textsubscript{rB} are the radiative rates of recombination, k\textsubscript{nrA} and k\textsubscript{nrB} are non-radiative rates of recombination, and k\textsubscript{TR} is the non-radiative electron transfer rate between states A and B. The non-radiative rates are temperature dependent and given by:    
\begin{center}
    $k_{nri} = \Gamma_{nri} e^{-E_{nri}/k_B T}$
    
    i = A, B, TR
\end{center}

Solving for n\textsubscript{A}(T) and under steady-state conditions gives Equation \ref{eq:nat}:

\begin{equation}
    n_A(T) = \frac{G_A(T)}{k_{rA} + k_{nrA} + k_{TR}}
    \label{eq:nat}
\end{equation}
\newline

We make the approximation that generation rates G\textsubscript{A}(T) and G\textsubscript{B}(T) are independent of temperature. Considering that the PL intensity from radiative A$\rightarrow$G transitions, I\textsubscript{A}(T), is equal to the radiative rate k\textsubscript{rA} times the excited state population n\textsubscript{A}(T), we obtain Equation \ref{eq:at}, where $G_A(0)=I_A(0)$:

\begin{equation}
    I_A(T) = I_A(0)\frac{k_{rA}}{k_{rA} + k_{nrA} + k_{TR}}
    \label{eq:at}
\end{equation}
\newline

When we write out the full, temperature-dependent forms of the non-radiative rates and re-arrange the result using the constants C\textsubscript{A} and C\textsubscript{TR} defined below, we obtain Equation \ref{eq:aexact}:

\begin{center}

   $C_A=\Gamma_{nrA}/k_{rA}$

    $ C_{TR}=\Gamma_{TR}/k_{rA}$ 

\end{center}

\begin{equation}
    I_A(T) = \frac{I_A(0)}{1 + C_Ae^{-E_{nrA}/k_BT} + C_{TR}e^{-E_{TR}/k_BT}}
    \label{eq:aexact}
\end{equation}
\newline

Solving for n\textsubscript{B}(T) under steady state conditions gives Equation \ref{eq:nbt}, which is dependent upon n\textsubscript{A}(T):

\begin{equation}
    n_B(T) = \frac{G_B(T) + n_A(T)k_{TR}}{k_{rB} + k_{nrB}}
    \label{eq:nbt}
\end{equation}

Again, multiplying $n_B(T)$ by the radiative rate $k_{rB}$ to get the PL intensity $I_B(T)$ and making the approximation that $G_B(T)$ is constant (such that $I_B(0)=G_B(0)$), we write Equation \ref{eq:bt}:

\begin{equation}
    I_B(T) = \frac{I_B(0)k_{rB} + n_A(T)k_{TR}k_{rB}}{k_{rB} + k_{nrB}}
    \label{eq:bt}
\end{equation}

Expanding equation \ref{eq:bt} gives us the following exact expression for n\textsubscript{B}(T):

\begin{equation}
    I_B(T) = I_B(0)\frac{k_{rB} }{k_{rB} + k_{nrB}} + (\frac{k_{TR}k_{rB}}{ k_{rB} + k_{nrB} })(\frac{I_A(0)}{k_{rA} + k_{nrA} + k_{TR}})
\end{equation}

We again use the proportional relationship between the PL intensity from radiative B$\rightarrow$G transitions and the population n\textsubscript{B}(T) to write an expression for I\textsubscript{B}(T). When we write out the full, temperature-dependent forms of the non-radiative rates and re-arrange the result using the constants C\textsubscript{A} and C\textsubscript{TR} as well as C\textsubscript{B} defined below, we obtain Equation \ref{eq:bexact}:

\begin{center}

 $ C_B=\Gamma_{nrB}/k_{rB}$ 
 
\end{center}

\begin{equation}
    I_B(T) = \frac{I_B(0)}{1 + C_Be^{-E_{nrB}/k_BT}} + \frac{ I_A(0)C_{TR} e^{-E_{TR}/k_BT}}{ (1 + C_Be^{-E_{nrB}/k_BT})(1 + C_Ae^{-E_{nrA}/k_BT} + C_{TR}e^{-E_{TR}/k_BT})}
    \label{eq:bexact}
\end{equation}

We now use Equations \ref{eq:aexact} and \ref{eq:bexact} to fit measured I(T) data, which correspond to the integral of the Gaussian fit for the R-Cu peak at every temperature, such that $I(T) = I_A(T) + I_B(T)$. We first use an approximate version of Equation \ref{eq:bexact} to fit the measured I(T) data, then take the resulting parameters as seed values for the fit to the exact equation. To write the approximate version of Equation \ref{eq:bexact}, we expand the product in the denominator of Equation \ref{eq:bexact}: 

\begin{center}
   $1 + C_Ae^{-E_{nrA}/k_BT} + C_Be^{-E_{nrB}/k_BT} + C_{TR}e^{-E_{TR}/k_BT} + C_BC_Ae^{-(E_{nrB}+E_{nrA})/k_BT } + C_BC_{TR}e^{-(E_{nrB}+E_{TR})/k_BT } $
\end{center}

This expanded product contains two terms with effective activation energies $E_{nrB}+E_{nrA}$ and $E_{nrB}+E_{TR}$. Assuming these effective activation energies are large compared to the measurement temperatures in our experiment, we neglect these terms in the approximate expression for I\textsubscript{B}(T). 

The total PL intensity is $I(0)=I_A(0)+I_B(0)$. We define a proportionality factor $W_A$ such that $W_A=I_A(0)/I(0)$ and $1-W_A=I_B(0)/I(0)$. The result is Equation \ref{eq:ibapprox}

\begin{equation}
        I_B(T) \approx \frac{(1-W_A)I(0)}{1 + C_Be^{-E_{nrB}/k_BT}} + \frac{ W_AI(0)C_{TR}e^{-E_{TR}/k_BT}}{ 1 + C_Ae^{-E_{nrA}/k_BT} + C_Be^{-E_{nrB}/k_BT} + C_{TR}e^{-E_{TR}/k_BT}}
        \label{eq:ibapprox}
\end{equation}

Equations \ref{eq:iaexactfit} and \ref{eq:ibexactfit} are the exact equations for $I_A(T)$ and $I_B(T)$ used to fit measured I(T) data, in terms of the fit parameters listed below.

\begin{equation}
            I_A(T) = \frac{W_AI(0)}{1 + C_Ae^{-E_{nrA}/k_BT} + C_{TR}e^{-E_{TR}/k_BT}}
        \label{eq:iaexactfit}
\end{equation}

\begin{equation}
        I_B(T) = \frac{(1-W_A)I(0)}{1 + C_Be^{-E_{nrB}/k_BT}} + \frac{ W_AI(0)C_{TR}e^{-E_{TR}/k_BT}}{ (1 + C_Be^{-E_{nrB}/k_BT})(1 + C_Ae^{-E_{nrA}/k_BT} + C_{TR}e^{-E_{TR}/k_BT})}
        \label{eq:ibexactfit}
\end{equation}
\newline

\newpage
We use the measurement results in Figure 6d to fix the value of W\textsubscript{a} for fitting and find that the energy parameters are reasonable well constrained while the coefficients C\textsubscript{A}, C\textsubscript{B}, and C\textsubscript{TR} are virtually unconstrained. The values of the parameters that best describe measured I(T) data are given below, with 68\% confidence intervals in parentheses: 

\begin{center}
    $W_A$ = 0.9285
    
    $I(0) = 1.074\times10^9$ counts (1.067, 1.08)
    
    $E_{nrA}$ = 105.8 meV (68.1, 143.5)
    
    $E_{nrB}$ = 214.3 meV (166.4, 262.1)
    
    $E_{TR}$ = 152.7 meV (131.8, 173.6)
    
    $C_A$ = 1372
    
    $C_B$ = 5301 
    
    $C_{TR} = 1.47\times10^4$

\end{center}

\clearpage
\newpage

\clearpage
\newpage

\section{10. Hybrid DFT Calculations for Formation Energies}

	\begin{figure}[!h]
      \begin{center}
          \includegraphics[width=12cm]{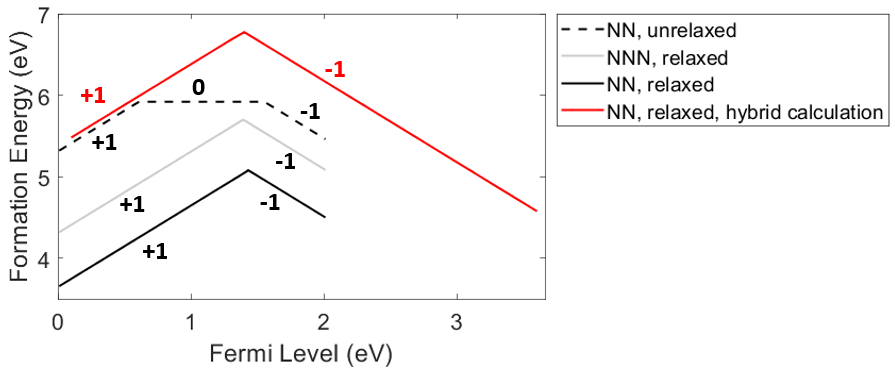}
              \caption{Formation energies for negatively charged, neutral, and positively charged (-1, 0, and +1 charges with respect to the ZnS lattice as indicated on plots) nearest- and next-nearest-neighbor (NN and NNN, respectively) associations of Cu\textsubscript{Zn} and V\textsubscript{S} in ZnS, as a function of the Fermi level. Solid lines indicate calculations performed for a relaxed lattice. Dashed lines indicate calculations performed for an unrelaxed lattice. Black and grey lines indicate DFT calculations, and the red line indicates hybrid DFT calculations. Details of the DFT settings are in the Methods section of the main text.}
          \label{table1}
      \end{center}
    \end{figure}

\clearpage
\newpage

\bibliography{Bib}